\documentclass[a4paper,11pt,nohyper]{JHEP3}
\usepackage{amsmath,latexsym,amssymb,slashed}
\usepackage{graphicx}
\usepackage{psfrag}
\usepackage[latin1]{inputenc}
\usepackage{subfigure}
\usepackage{slashed}
\usepackage{tensor}
\usepackage{bbm}

\newcommand\beal{\begin{align}}
\newcommand\nn{\nonumber}
\newcommand\bbone{\ensuremath{\mathbbm{1}}}

\newcommand{\li}{l_{\mathrm{int}}}

\newcommand{\eq}[1]{\begin{equation}#1\end{equation}}
\newcommand{\spl}[1]{\begin{split}#1\end{split}}

\newcommand{\mcal}{\mathcal{M}}

\newcommand{\ncal}{\mathcal{N}}
\newcommand{\ocal}{\mathcal{O}}

\newcommand{\G}{\Gamma}
\newcommand{\g}{\gamma}
\newcommand{\e}{\epsilon}
\newcommand{\we}{\widetilde{\eta}}

\renewcommand{\O}{\Omega}

\newcommand{\boxedeq}[1]{
\begin{equation}
\fbox{
\rule[0.7cm]{0pt}{0pt}
$#1$
\rule[-0.45cm]{0pt}{0pt}
}
\end{equation}
}

\def\d{\text{d}}

\def\slashchar#1{\setbox0=\hbox{$#1$}           
\dimen0=\wd0                                 
\setbox1=\hbox{/} \dimen1=\wd1               
\ifdim\dimen0>\dimen1                        
\rlap{\hbox to \dimen0{\hfil/\hfil}}      
#1                                        
\else                                        
\rlap{\hbox to \dimen1{\hfil$#1$\hfil}}   
/                                         
\fi}

\usepackage{xcolor}
\usepackage{color}

\newcommand{\cnote}[1]{}

\title{3d \boldmath$\ncal=1$\boldmath{} effective supergravity  and F-theory\\ from M-theory on fourfolds}

\author{Dani\"{e}l Prins and Dimitrios Tsimpis\\
Universit\'{e} de Lyon\\
UMR 5822, CNRS/IN2P3, Institut de Physique Nucl\'{e}aire de Lyon\\
4 rue Enrico Fermi,
F-69622 Villeurbanne Cedex,  France\\

E-mail:
\email{dlaprins@ipnl.in2p3.fr}, \email{tsimpis@ipnl.in2p3.fr}
}

\abstract{We consider 3d $\ncal=1$ M-theory compactifications on Calabi-Yau fourfolds, and the effective 3d theory of light modes obtained by reduction from eleven dimensions. 
We study in detail the mass spectrum at the vacuum and, by decoupling the massive multiplets, we derive the effective 3d $\ncal=1$ theory in the large-volume limit up to quartic fermion terms. We show that in general it is an ungauged $\ncal=1$ supergravity of the form expected from 3d supersymmetry. 
In particular the massless bosonic fields 
consist of the volume modulus and the axions originating from the eleven-dimensional three-form, while the moduli-space metric is locally isometric to hyperbolic space. We consider  the F-theory interpretation of the 3d $\ncal=1$ M-theory vacua  
in the light of the F-theory effective action approach. We show that these vacua generally have F-theory duals with circle fluxes, thus breaking 4d Poincar\'e{} invariance. 
}

\begin{document}
\setcounter{footnote}{0}
\renewcommand{\thefootnote}{\arabic{footnote}}
\setcounter{section}{0}
\vfill\break
\section{Introduction}\label{introduction}\label{sec1}

Perturbative vacua of string theory can be constructed by extremizing the low-energy effective action (leea), i.e. the generating functional of one-particle irreducible Green's functions for the ten-dimensional massless fields. The leea can be constructed systematically  to any desired order in the string coupling and in $\alpha'$, at least in principle, either by sigma-model perturbation or by scattering amplitude methods. 
In the large-volume limit of M-theory compactifications the analog of the 
 $\alpha'$ expansion is an expansion in powers of the Planck length $l_P$ which, in units where the compactification radius is equal to one, becomes a small dimensionless parameter. 

In M-theory the leea cannot be constructed systematically from first principles, because of the lack of a complete microscopic formulation of the theory, but it can by argued that anomaly cancelation together with supersymmetry should suffice to uniquely constrain the leading-order correction  which occurs at order $l_P^6$ \cite{Howe:2003cy}. Moreover in certain cases a general inductive argument based on supersymmetry can be formulated to show the existence of supersymmetric vacua of M-theory to any finite order in 
$l_P$ \cite{Becker:2014rea}. 

3d M-theory flux vacua with $\ncal=2$ supersymmetry (four real  supercharges) from compactification on Calabi-Yau (CY) fourfolds were constructed in \cite{Becker:1996gj}. They were shown to solve the eleven-dimensional equations of motion of M-theory to order $l_P^3$ in \cite{Becker:2001pm}, as recently reviewed in \cite{Grimm:2014xva} in a scheme where the 11d equations of motion are solved perturbatively in an expansion in powers of $l_P$. The four-form flux in these vacua is ``small'' as it is quantized in units of $l_P^3$. Interestingly  the tadpole cancelation constraint that the flux in these vacua has to satisfy arises at order $l_P^6$ in the equations of motion, whereas the flux itself is  $\mathcal{O}(l_P^3)$.

The vacua of \cite{Becker:1996gj} constitute the starting point of many F-theory constructions \cite{Vafa:1996xn}, see \cite{Denef:2008wq} for a review. Although this has not been proven to date, it is believed that these vacua should survive to all orders in $l_P$. More precisely this means that to any finite order in $l_P$ the metric of the fourfold can be corrected so that $\ncal=2$ supersymmetry is preserved to that order. This has recently been partially verified in \cite{Grimm:2014xva} where it was shown that the metric gets corrected at order $l^6_P$ away from the Ricci-flat CY metric while still retaining its K\"{a}hler property.

In \cite{Prins:2013wza} it was shown that some of the conditions of \cite{Becker:1996gj} can be relaxed in order to obtain three-dimensional $\ncal=1$ vacua on CY fourfolds solving the M-theory equations of motion to order $l^3_P$. 
These vacua can also be seen from the point of view of the effective three-dimensional theory arising from M-theory compactification on CY fourfolds: $\ncal=2$ gauged supergravity in three dimensions \cite{Haack:1999zv,Haack:2001jz}. From the point of view of the effective 3d theory they arise as partially supersymmetry-breaking vacua {\it spontaneously} breaking supersymmetry from $\ncal=2$ to $\ncal=1$ \cite{Berg:2002es}.

Given the central role of the $\ncal=2$ vacua for F-theory it is important to study the properties of their $\ncal=1$ spontaneously-broken counterparts in more detail. In the present paper we perform a Kaluza-Klein (KK) reduction of the 
11d theory around the $\ncal=1$ vacuum of \cite{Prins:2013wza} up to  quartic fermion terms. It is interesting to note that the quadratic fermion terms are linear in the flux and are thus of order $l_P^3$; they capture the superpotential and the ``real superpotential'' of 3d gauged   $\ncal=2$  supergravity, both of which are linear in the flux and hence are visible at order $l_P^3$. With this information one  can then compute the potential of the effective 3d theory {\it exactly} to order $l_P^6$: since the latter is quadratic in the superpotentials, one does not need to know the   $\mathcal{O}(l_P^6)$ corrections to the superpotentials or the K\"{a}hler potential in order to compute the potential to order $l_P^6$.

Moreover at the vacuum some of the fields will generally obtain masses. These are linear in the flux and are hence of order $l_P^3$. Contrary to  the 
bosonic mass terms which enter quadratically in the 3d action and are thus  of order $l_P^6$, the fermionic mass terms enter linearly and are of order $l_P^3$. 
I.e., if one only looks at the bosonic part of the action, one needs the   $\mathcal{O}(l_P^6)$ terms in order to read off the masses. However this same mass spectrum is already accessible at order $l^3_P$, provided one looks at the quadratic fermion terms.

A related observation is the following. In a KK reduction around a fluxless vacuum, the light modes of the 3d effective theory can be obtained by giving 3d spacetime dependence to the parameters (moduli) of the 11d solution. 
Turning on flux results in some of the moduli being lifted. Since their masses at the vacuum are linear in the flux and are 
thus of order $l_P^3$, they can be taken to be much smaller than the KK scale. To order $l_P^3$ it is therefore justified to keep the {same} light mode expansion in the case with non-vanishing four-flux as in the fluxless case. The expansions for the light modes will generally get modified at order $l_P^6$, where the 
fourfold geometry gets corrected away from Ricci-flatness \cite{Grimm:2014xva,Grimm:2014efa}. However these modifications are subleading, as far as the mass spectrum of the light modes is concerned, and will not be necessary for our analysis.\footnote{The observations of the 
three previous paragraphs   
might be relevant in understanding the nontrivial cancellations observed in \cite{Grimm:2014efa} among some of the terms in the 3d $\ncal=2$ effective action at order $l_P^6$.}

In the present paper we study in detail the mass spectrum of light fields of the 3d effective theory at the $\ncal=1$ vacuum. In particular we show how the original $\ncal=2$ massless supermultiplets get reorganized in terms of $\ncal=1$ massive and massless supermultiplets. By decoupling the massive supermultiplets we derive  the $\ncal=1$ effective theory below the partial supersymmetry-breaking scale up to quartic fermion terms. We show that it is an ungauged $\ncal=1$ supergravity of the general form expected by 3d supersymmetry \cite{deWit:2003ja,deWit:2004yr}. In particular the massless bosonic fields below the partial supersymmetry-breaking scale  
consist of the volume modulus and the ``axions'' originating from the eleven-dimensional  three-form. Moreover the moduli-space metric is locally isometric to hyperbolic space.

We next consider the question of the F-theory interpretation of the  $\ncal=1$ vacua of \cite{Prins:2013wza}. As we review in the following, a dictionary was put forward in  \cite{Grimm:2010ks,Grimm:2011sk} relating the 3d $\ncal=2$ gauged supergravities arising from M-theory compactifications on elliptically-fibered CY fourfolds to 4d $\ncal=1$ effective actions from F-theory CY compactifications. In particular this relation imposes certain constraints on the four-form flux which should be obeyed for the M-theory compactification to admit an F-theory interpretation. 
We show that M-theory CY compactification down to $\mathbb{R}^{1,2}$ with four-flux as in the  $\ncal=1$ vacua of \cite{Prins:2013wza} is dual to F-theory compactification down to $\mathbb{R}^{1,2}\times S^1$. Moreover the latter breaks 4d Poincar\'e{} invariance, even in the large radius limit, due to the presence of quantized ``circle fluxes'' threading the $S^1$. The question of whether or not the F-theory dual preserves 4d Poincar\'e{} invariance translates on the M-theory side to the question of whether or not the four-form has exactly one leg on the elliptic fiber \cite{Dasgupta:1999ss},\footnote{The precise statement is that the integral of the four-form over any divisor of the base $B$ of the elliptically-fibered CY, and the integral of the four-form over the elliptic fibration of any holomorphic curve in $B$ must vanish.} so the presence of the 
circle fluxes can also be understood from this geometric perspective.

The outline of the remainder of the paper is as follows. In section \ref{sec:sol} we review the 3d $\ncal=1$ M-theory vacua of \cite{Prins:2013wza}. Section \ref{sec:eff} first reviews the 3d $\ncal=2$ effective supergravity obtained from the KK reduction of M-theory in the presence of flux. Section \ref{sec:vacm} then analyzes the spectrum of masses at the $\ncal=1$ spontaneously broken vacuum and reviews the different super-Higgs mechanisms. The 3d $\ncal=1$ effective supergravity below the partial supersymmetry-breaking scale is given in section \ref{sec:eff3d}. In section \ref{sec:ftheory} we first review the F-theory effective action approach of \cite{Grimm:2010ks,Grimm:2011sk} before giving the F-theory interpretation of the  $\ncal=1$ M-theory vacuum of \cite{Prins:2013wza}. 
We conclude in section \ref{sec:con}. 
To keep the presentation simple we have moved most technical details to the four appendices.

\section{Three-dimensional \boldmath$\ncal=1$~\unboldmath{}solutions from CY fourfolds}\label{sec:sol}

Let us review the 3d $\ncal=1$ vacua arising from M-theory large-volume compactification on CY fourfolds \cite{Prins:2013wza}. 
The perturbative expansion in $l_P$ is justified provided $l_P$ is small compared to some other length scale. The latter will be taken to be the `radius', $\li$, of the internal CY manifold at the vacuum, which we define by:
\eq{\label{vacv}\li:=\mathring{V}^{\frac18}~,}
where $\mathring{V}$ is the volume of the vacuum CY.\footnote{A note on notation: the circle over $V$ is used to distinguish the moduli-dependent volume, $V$, from the value of the latter at the vacuum, $\mathring{V}$. This distinction will become important in the following, cf. section \ref{sec:eff}.}
We will henceforth work with units where $\li=1$, so that $l_P$ is dimensionless and satisfies $l_P<<1$.

The eleven-dimensional metric reads: 
\eq{\d s^2=\eta_{\mu\nu}\d x^{\mu}\d x^{\nu}+\mathring{g}_{mn}\d y^m\d y^n+\mathcal{O}(l_P^6)~,}
where $\mathring{g}_{mn}$ is a CY metric; $x^{\mu}$, $\mu=0,1,2$, and $y^m$, $m=1,\dots,8$, are 3d spacetime and internal CY coordinates respectively.

The four-flux reads:
\eq{G_{mnrs}=\mathring{G}_{mnrs}+\mathcal{O}(l^6_P)
~;~~~
G_{\mu\nu\rho s}=\mathcal{O}(l^6_P)
~,}
where
\eq{\label{vfn1}
\mathring{G}=g \left(\mathring{J} \wedge \mathring{J}+\frac{6}{|\mathring{\Omega}|}~\!\mathrm{Re}\mathring{\O} \right)
+g^{(2,2)}
~,}
with $g$ an order-$l^3_P$ real constant and $g^{(2,2)}$ an order-$l^3_P$ (2,2)-primitive four-form; $(\mathring{J}$, $\mathring{\Omega})$ denote the K\"{a}hler form and holomorphic four-form of the vacuum CY respectively.\footnote{The different normalization with respect to (3.19) of \cite{Prins:2013wza} is due to the fact that $\Omega$ therein is identified with $\hat{\Omega}$ of the present paper, cf., (\ref{b5}).} 
Setting $g=0$ reduces to the $\ncal=2$ vacuum of \cite{Becker:1996gj}.

The equations of motion to order $l_P^3$ imply that $G=\mathring{G}$ must be harmonic,
\eq{\d G=\d*_8 G=0~,
}
and, in the presence of $N_{\mathrm{M2}}$ M2-branes, satisfies the tadpole cancellation condition,
\eq{
\frac{1}{2(2\pi l_P)^6}\int_{\mathcal{M}_8}
G\wedge G+N_{M2}=\frac{\chi(\mathcal{M}_8)}{24}
~,}
where $\chi(\mathcal{M}_8)$ is the Euler character of the fourfold. 
We will henceforth set $N_{\mathrm{M2}}=0$.

In addition the four-form flux is subject to the quantization condition  \cite{wq}:
\eq{\label{wqc}
\frac{[G]}{(2\pi l_P)^3}-\frac{c_2}{2}\in H^4(\mcal_8,\mathbb{Z})
~,}
where $[G]$ is the cohomology class of $G$. Let us expand the four-form at the vacuum as follows:
\eq{\label{ge}G=\sum_{i=1}^{b_4}c^i\omega_i
~,}
where $\omega_i$, $i=1,\dots,b_4$, are harmonic four-forms on the vacuum CY so that the $[\omega_i]$'s form a basis of the integral cohomology $H^4(\mcal_8,\mathbb{Z})$, and we have taken into account that $G=\mathring{G}$ is harmonic. Furthermore  
let $\mathcal{C}_i$, $i=1,\dots,b_4$, be a basis of $H_4(\mcal_8,\mathbb{Z})$, so that $\mathcal{C}_i$ is Poincaré dual to $[\omega_i]$. 
The wedge product of the $\omega_i$'s is normalized with respect to the 
volume of the vacuum CY, which we take to be equal to one as mentioned below (\ref{vacv}):
\eq{\int_{\mathcal{M}_8}\omega_i\wedge\omega_j=
\int_{\mathcal{C}_i}\omega_j
=\int_{\mathcal{C}_j}\omega_i=N_{ij}
~,}
where 
$N_{ij}\in\mathbb{Z}$ is the oriented intersection number $\#(\mathcal{C}_i,\mathcal{C}_j)$. 
The quantization condition (\ref{wqc}) then imposes: 
\eq{c^i={(2\pi l_P)^3}N^i
~,}
where the $N^i$'s can be integer or half-integer. 
Thus the flux is quantized in units of $l_P^3$ in accordance with 
claim that the flux is small.

\section{Three-dimensional effective action}\label{sec:eff}

In the case without flux ($G=0$) let us collectively denote the bosonic parameters (moduli) of the vacuum solution, to be described in more detail in \ref{sec:expb}, by 
\eq{\Phi:=(M^A, Z^{\alpha}, \mathcal{A}^A_{\mu}, N^I)~,}
where $A=1,\dots, h^{1,1}$; $\alpha=1,\dots, h^{3,1}$; $I,\dots, h^{2,1}$, and 
$h^{p,q}$ are the Hodge numbers of the CY fourfold. The 
$M^A$ are real scalars parameterizing K\"{a}hler deformations, while the $Z^{\alpha}$ are complex scalars parameterizing complex structure deformations of the CY, cf.~(\ref{47}). The  $N^I$ are complex scalars (axions) parameterizing a torus $H^{2,1}(\mcal_8)/H^3(\mcal_8,\mathbb{Z})$ of complex dimension $h^{2,1}$, while $\mathcal{A}^A_{\mu}$ are 3d vectors; both originate from the eleven-dimensional three-form, cf.~(\ref{cexp}).

These parameters are the coordinates of the moduli space of solutions to the 
equations of motion of eleven-dimensional supergravity. 
In order to obtain the 3d effective action capturing the physics of small fluctuations around the vacuum, defined to correspond to the point $\mathring{\Phi}$ in the moduli space, we  
promote the moduli variations $\delta{\Phi}$ to 3d spacetime fields $\delta{\Phi}(x)$ so that:
\eq{\label{pro}\Phi\longrightarrow \Phi(x):=\mathring{\Phi}+\delta{\Phi}(x)~.}

The procedure described above is equivalent to a KK reduction with the infinite KK tower truncated to the massless level. 
Turning on the four-flux ($G\neq0$) results in some of the moduli obtaining masses at the vacuum, which are linear in the flux and are therefore of order $l_P^3$. Since these masses can be taken to be much smaller than the KK scale, it is justified to keep the {same} light mode expansion in the case with non-vanishing four-flux as in the fluxless case. Althought the expansions for the light modes will generally get modified at order $\mathcal{O}(l_P^6)$ \cite{Grimm:2014xva,Grimm:2014efa}, these modifications are subleading as far as the mass spectrum is concernced and will not be necessary for our analysis.

We then insert the expansions (\ref{kk1}) in the eleven-dimensional action\footnote{Supersymmetry alone allows for a
supersymmetric correction to eleven-dimensional supergravity already at order $\ocal(l_P^3)$ \cite{Tsimpis:2004rs}.}
\eq{\spl{\label{s}S=S^{\mathrm{b}}+S^{\mathrm{f}}+\ocal(l_P^6)
~,}}
where the bosonic and fermionic parts of the action are given respectively by \cite{Cremmer:1978km},
\eq{\label{sb}S^{\mathrm{b}}=\frac{1}{2\kappa^2}\int\d^{11}x\sqrt{-g}
\big(R(g)-\frac{1}{2}G^2 \big)
-\frac{1}{12\kappa^2}\int C\wedge G\wedge G
~,}
and 
\eq{\spl{\label{sf}S^{\mathrm{f}}=\frac{1}{2\kappa^2}\int\d^{11}x\sqrt{-g}&\big[
2\tilde{\psi}_{M}\Gamma^{MNR}\nabla_N\psi_R\\
-&\frac{1}{48}
(\tilde{\psi}_M\Gamma^{MNPQRS}\psi_N+12\tilde{\psi}^P\Gamma^{QR}\psi^S)G_{PQRS}
 +\cdots\big]
~,}}
where $g_{MN}$, $G_{MNPQ}$ and $\psi_M$ are the eleven-dimensional metric,  four-form flux and gravitino respectively; 
the ellipses in (\ref{sf}) denote the quartic fermion terms, whose explicit form will not be necessary in the following.

Next we keep only up to terms quadratic in the fluctuations. For this last step it is important to 
note that $\mathring{\Phi}$ does not depend on the  spacetime coordinates, so that $\partial_{\mu}\Phi(x)$ is linear in the 
variations. Finally integrating over the internal CY coordinates, 
and after certain standard manipulations described in more detail in appendix \ref{app:reduction}, one obtains the 
bosonic part of the 3d effective action (\ref{sb3d}), or equivalently \cite{Haack:1999zv,Haack:2001jz}:
\eq{\spl{\label{sb3d2}
S^b=\frac{1}{2\kappa^2}\int\d^3x \Big\{\sqrt{-g_{(3)}}\Big[
{R}&-\hat{G}_{AB}(\partial_{\nu}\hat{M}^A\partial^{\nu}\hat{M}^B+\frac12  F_{\mu\nu}^{A}F^{B\mu\nu})
\\
&-2G_{\alpha\bar{\beta}}
\partial_{\nu}Z^{\alpha}\partial^{\nu}\bar{Z}^{\beta}
-2\hat{G}_{I\bar{J}}D_{\mu}N^ID^{\mu}\bar{N}^J\Big]\\
&-\frac{1}{2}\varepsilon^{\mu\nu\rho}d_{AI\bar{J}}\mathcal{A}^A_{\mu}
D_{\nu}N^ID_{\rho}\bar{N}^J
+\varepsilon^{\mu\nu\rho}\Theta_{AB}\mathcal{A}_{\mu}^A F^B_{\nu\rho} 
+\mathcal{O}(l_P^6)\Big\}~,}}
where the various sigma-model couplings are defined in appendix \ref{app:reduction}; $\hat{M}^A$ is related to ${M}^A$ via (\ref{kred}). Note that the last Chern-Simons term is linear in the flux and hence of order $l^3_P$; all other terms are $\mathcal{O}(1)$.

The bosonic moduli are paired up with their fermionic superpartners, discussed in more detail in section \ref{sec:expf}, to form 3d massless $\ncal=2$ supergravity multiplets:  
\boxedeq{
\mathrm{gravity:}~\big(g_{\mu\nu};\chi_{\mu}\big)~;~~~
\mathrm{vector:}~\big(\mathcal{A}^{A},M^A;\lambda^{A}\big)~;~~~
\mathrm{scalar:}~\big(Z^{\alpha};\lambda^{\alpha}\big)~,
~\big(N^I;\lambda^{I}\big)~,\nn
}
where $\chi_{\mu}$ is a complex gravitino and $\lambda^{A}$, $\lambda^{\alpha}$, $\lambda^{I}$ are complex 3d spinors. 
All multiplets contain 2+2 real bosonic and fermionic physical degrees of freedom except for the gravity multiplet which contains none. 

At the vacuum some of the fields will generally obtain masses. 
It is interesting to note that the fermionic mass terms enter linearly in the action, contrary to the bosonic mass terms which enter quadratically. 
Moreover, as mentioned earlier, the masses of the light modes are of order $l_P^3$ hence the fermionic mass terms in the 3d action are order $l_P^3$ whereas the bosonic mass terms, which are quadratic in the mass, are of order $l_P^6$. Therefore one does {\it not} need to know the $\mathcal{O}(l_P^6)$ terms in the 3d action in order to read off the mass spectrum. We will make use of this observation in section \ref{sec:vacm}.

\subsection{Masses at the $\ncal=1$ vacuum}\label{sec:vacm}

For the purposes of this section, and this section only, we will fix all bosonic moduli to their vacuum values: $\Phi\rightarrow\mathring{\Phi}$. We will however omit the circles above the bosonic fields to keep the notation simple.

The direction of $J$ inside $H^2(\mcal_8,\mathbb{R})$ defines a vector $M^A$, cf.~(\ref{411}), while $V_A$, defined in (\ref{d4}), can be thought of as its dual in view of (\ref{mb}). The projector $R_A{}^B$, defined in (\ref{rd}), projects on the left onto the vertical space of $M^A$ while on the right it projects onto the vertical space of $V_A$, cf.~(\ref{rprop}). 
Furthermore the projector $R_A{}^B$ projects onto the direction  of the primitive  cohomology (with respect to $J$ of the CY):
\eq{\label{lr1}{e}_{A}':={R}_{A}{}^{B}e_B~;~~~
M^A{e}_{A}'=0
~,}
where  the second equation follows from (\ref{rprop}); ${e}_{A}'$ defined above is indeed the primitive part of $e_A$, $J\lrcorner e_{A}'=0$, as can be seen from (\ref{dec2}),(\ref{411}),(\ref{d6}):
\eq{\label{pc}
e_A=e_{A}'+\frac{V_A}{V}J
~.}
Note in particular that there are $(h^{1,1}-1)$ independent $e_{A}'$'s, since there is one linear relation among them.

Moreover we can use the projector to split the vector multiplets into horizontal and vertical directions. Explicitly for the fermions we define:
\eq{\label{lr2}
\lambda:=\frac{V_A}{V}{\lambda}^{A}
~;~~~
{\lambda}^{\prime B}:=\lambda^A R_A{}^{B}=\lambda^A-\lambda M^A
~;~~~V_A{\lambda}^{\prime A}=0
~,}
where the last equation follows from (\ref{rprop}). There are thus $(h^{1,1}-1)$ independent ${\lambda}^{\prime A}$'s such that:
\eq{\label{lexp}
\lambda^Ae_A={\lambda}^{\prime A}{e}_{A}'+\lambda J
~.}
Similarly for the vectors we define,
\eq{\label{va}\mathcal{A}:=\frac{V_A}{V}\mathcal{A}^A~;~~~
\mathcal{A}^{\prime A}:=\mathcal{A}^{B}R_B{}^{A}=\mathcal{A}^{A}
-\mathcal{A}M^{A}~;~~~V_A\mathcal{A}^{\prime A}=0
~,}
such that, in analogy to (\ref{lexp}), 
\eq{\label{lexpa}
\mathcal{A}^A\wedge e_A=\mathcal{A}^{\prime A}\wedge{e}_{A}'
+\mathcal{A}\wedge J
~.}
Moreover for 
the field-strengths we set: $F:=\d\mathcal{A}$, 
$F^{\prime A}:=\d\mathcal{A}^{\prime A}$.

\subsection*{The super-Higgs mechanism}

The super-Higgs mechanism by which the gravitino and vector $\ncal=1$ multiplets obtain mass has been described in \cite{Hohm:2004rc}. 
The massive $\ncal=1$ gravitino multiplet consists of a massive gravitino and a massive vector. In our case the massive gravitino results from $\chi^-_{\mu}$ eating the spinor field $\lambda^+$. This can be seen by examining the quadratic fermion terms at the $\ncal=1$ Minkowski vacuum, cf.~(\ref{j5}), (\ref{i6}),
\eq{\spl{\label{l1}
S=\int\d^3x~V\big[&3^22^3
(\tilde{\lambda}^{+}\gamma^{\nu}\partial_{\nu}{\lambda}^{+})
-(\tilde{\chi}^{\prime-}_{\mu}\gamma^{\mu\nu\rho}\partial_{\nu}
\chi^{\prime-}_{\rho})\\
-&6g(\tilde{\chi}_{\mu}^{\prime-}\gamma^{\mu\nu}\chi_{\nu}^{\prime-})
-2^43^2ig(\tilde{\lambda}^{+}\gamma^{\nu}\chi_{\nu}^{\prime-})
-2^43^4g
(\tilde{\lambda}^{+}\lambda^{+})+~\!\cdots\big]
~,}}
where the ellipses denote terms which do not contain $\chi^-_{\mu}$ or $\lambda^+$; the volume factor comes from integrating over the CY coordinates. 
It can be checked that all dependence on $\lambda^+$ can be absorbed by redefining:
\eq{
\hat{\chi}_{\mu}
=\chi_{\mu}^{\prime-}+\frac{1}{g}\partial_{\mu}\lambda^+-6\gamma_{\mu}\lambda^+
~,}
so that in terms of $\hat{\chi}_{\mu}$ the Lagrangian (\ref{l1}) takes the form:
\eq{\label{l2}
S=\int\d^3x~V\big[(\hat{\chi}_{\mu}\gamma^{\mu\nu\rho}\partial_{\nu}
\hat{\chi}_{\rho})+6g(\hat{\chi}_{\mu}\gamma^{\mu\nu}\hat{\chi}_{\nu})
+~\!\cdots\big]
~.
}
The volume factor can be reabsorded by performing the same Weyl rescaling as in (\ref{wr}), $g_{\mu\nu}\rightarrow V^{-2}g_{\mu\nu}$, together with a rescaling of the fermions, $\hat{\chi}_{\mu}\rightarrow V^{-\frac12}\hat{\chi}_{\mu}$, upon which (\ref{l2}) becomes:
\eq{\label{l3}
S=\int\d^3x~\!\big[(\hat{\chi}_{\mu}\gamma^{\mu\nu\rho}\partial_{\nu}
\hat{\chi}_{\rho})+6gV^{-1}(\hat{\chi}_{\mu}\gamma^{\mu\nu}\hat{\chi}_{\nu})
+~\!\cdots\big]
~.
}
It follows from the resulting equation of motion that  the Lagrangian above describes a gravitino of mass $6gV^{-1}$. Recall that a massless gravitino in three spacetime dimensions does not carry any degrees of freedom. Thanks to the Higgs mechanism described above,  the originally massless topological (non-propagating) gravitino becomes propagating by eating the one degree of freedom of the Goldstone fermion $\lambda^+$.

\subsection*{The topological Higgs mechanism}

A similar situation occurs for the massless vectors, which carry no degrees of freedom in three dimensions.  The Higgs mechanism in this case is somewhat unusual: it is topological, in the sense that the vectors become massive not by eating a Goldston scalar but rather by virtue of their Chern-Simons couplings \cite{Schonfeld:1980kb,Deser:1981wh}. The relevant terms in the Lagrangian read:
\eq{
S=\int\d^3x~\!\big(
-\frac12 V^2G_{AB} F_{\mu\nu}^{A}F^{B\mu\nu}
+\varepsilon^{\mu\nu\rho}\Theta_{AB}\mathcal{A}_{\mu}^A F^B_{\nu\rho} 
+\cdots\big)
~,}
where we have restricted the scalars to their values at the $\ncal=1$ Minkowski vacuum and have omitted 
terms which do not depend on $\mathcal{A}_{\mu}^A$. Evaluating the matrix $\Theta_{AB}$ at the vacuum taking (\ref{td}), (\ref{vfn1}) into account we obtain:
\eq{\Theta_{AB}=\Theta_{AB}'
-g V \big(G'_{AB}-6\frac{V_AV_B}{V^2}\big)~
~,}
where we defined, 
\eq{\spl{\label{319}
\Theta_{AB}'&:=\frac14\int g^{(2,2)}\wedge e_A\wedge e_B=\frac{V}{16}~g_{mnpq}^{(2,2)}~\!
e^{\prime mn}_A e^{\prime pq}_B
\\
G'_{AB}&:=R_{A}{}^{C}R_{B}{}^{D}G_{CD}=
G_{AB}-2\frac{V_AV_B}{V^2}
~,}}
and we used (\ref{d4}),(\ref{d6}),(\ref{c10}),(\ref{ma}); 
the second equality in the first line above follows from the fact that $g^{(2,2)}$ is primitive, and the integrand is a harmonic top form hence equal to a constant times the volume form. Note that $G'_{AB}$ has rank $(h^{1,1}-1)$, as follows from the properties of the projector $R_{A}{}^{B}$, cf.,~(\ref{rprop}). Inserting the explicit form of $\Theta_{AB}$ into the Lagrangian we obtain,
\eq{\spl{\label{l4}
S=\int\d^3x~\!{V}^2\big(
-&F_{\mu\nu}F^{\mu\nu} 
+6 gV^{-1}\varepsilon^{\mu\nu\rho}\mathcal{A}_{\mu} F_{\nu\rho} 
\\
-&\frac12 G'_{AB} F_{\mu\nu}^{\prime A}F^{\prime B\mu\nu}
+V^{-2} (
\Theta'_{AB}-gVG'_{AB}
)\varepsilon^{\mu\nu\rho}\mathcal{A}_{\mu}^{\prime A} F^{\prime B}_{\nu\rho} 
+\cdots\big)
~,}}
where we have taken (\ref{va}) into account. 
From the first line of 
(\ref{l4}) we see that the vector $\mathcal{A}$ obtains a mass equal to 
$6gV^{-1}$, i.e. degenerate with the mass of the gravitino $\hat{\chi}_{\mu}$ 
discussed previously.\footnote{In three spacetime dimensions  a vector of mass $m$ can be described by a Lagrangian of the form $\mathcal{L}=F_{\mu\nu}F^{\mu\nu}-m\varepsilon^{\mu\nu\rho}\mathcal{A}_{\mu} F_{\nu\rho}$.} From the second line of (\ref{l4}) we see that the vectors $\mathcal{A}^{\prime A}$ will also obtain masses in general; the number of massive vectors among the $\mathcal{A}^{\prime A}$'s will be equal to the rank of the matrix in the parenthesis in the last line of (\ref{l4}): assuming $G'_{AB}$ is invertible and $g^{(2,2)}$ is generic, this matrix will have maximal rank equal to $(h^{1,1}-1)$.

More specifically, as can be seen from (\ref{l4}), the masses of the  $\mathcal{A}^{\prime A}$'s are given by the eigenvalues of the matrix
\eq{
2 gV^{-1}(\bbone - \frac{1}{gV}(S^{-1})^T\Theta S^{-1})
~,}
where we have taken into account  
that we can put $G'$ in diagonal form: $G'=S^TS$ for some invertible matrix $S$ of rank  $(h^{1,1}-1)$.  
The upshot is that the Chern-Simons terms will generally give masses to all vectors and there is always one vector whose mass is degenerate with that of the massive gravitino:
\eq{m_{\mathcal{A}}= 6 gV^{-1}=m_{\hat{\chi}}~.}
Moreover from (\ref{j5}),(\ref{i6}),(\ref{319}) we obtain the following quadratic fermion terms at the $\ncal=1$ Minkowski vacuum, 
\eq{\spl{\label{l6}
S=\int\d^3x\sqrt{-g_{(3)}}~\Big(&\frac12 G'_{AB}[
(\tilde{\lambda}^{\prime A+}\gamma^{\nu}\nabla_{\nu}{\lambda}^{\prime B+})
+(\tilde{\lambda}^{\prime A-}\gamma^{\nu}\nabla_{\nu}{\lambda}^{\prime B-})]\\
-&
V^{-2}(
\Theta'_{AB}-gVG'_{AB}
)(\tilde{\lambda}^{\prime A+}{\lambda}^{\prime B+})\\
-&
V^{-2}(
\Theta'_{AB}+2gVG'_{AB}
)(\tilde{\lambda}^{\prime A-}{\lambda}^{\prime B-})
+~\!\cdots\Big)
~,}}
where the ellipses denote terms which do not contain $\lambda^{\prime A}$; we have also performed the same Weyl rescaling as in (\ref{wr}), $g_{\mu\nu}\rightarrow V^{-2}g_{\mu\nu}$, together with a rescaling of the fermions, $\lambda^{\prime A}\rightarrow 2^{-\frac32}V^{\frac12}\lambda^{\prime A}$ in order to get canonical kinetic terms. Reasoning as before, we see that for generic 
$g^{(2,2)}$ and invertible $G'_{AB}$ all the  $\lambda^{\prime A\pm}$'s get 
masses at the $\ncal=1$ vacuum.\footnote{
Even with the partial supersymmetry-breaking parameter switched off, $g=0$, 
it follows from (\ref{l6}) that  K\"{a}hler moduli other than the volume modulus may generally get masses depending on the 
form of $\Theta'_{AB}$. Anticipating the F-theory interpretation let us mention that for  
compactifications on smooth, full $SU(4)$-holonomy CY fourfolds and four-fluxes that respect 4d Poincar\'e{} invariance in F-theory, $\Theta'_{AB}$ vanishes 
identically (cf. the last paragraph of section \ref{sec:ftheory}). This is in accordance with the fact that in this case the potential for the K\"{a}hler moduli is flat at the classical level, see e.g. \cite{Denef:2008wq}. 
} 
It can also be seen, by comparing (\ref{l4}) and (\ref{l6}), that the masses of the $\lambda^{\prime A+}$'s  are degenerate with those of the $(h^{1,1}-1)$ vectors $\mathcal{A}^{\prime A}$:
\eq{m_{\mathcal{A}'}= m_{\lambda^{\prime+}}~.}
The mass terms of the complex scalar moduli can also be read off from (\ref{j4}),(\ref{i5}), giving a contribution to the 3d Lagrangian proportional to:
\eq{\label{jj5}\mathcal{L}\sim
(\bar{\Phi}_{\bar{\alpha}}\cdot\Phi_{\beta})\big[
(\tilde{\lambda}^{\alpha c}\gamma^{\nu}\nabla_{\nu}{\lambda}^{\beta})
+3g
(\tilde{\lambda}^{\alpha c}\lambda^{\beta})\big]
+\frac{1}{128}\Phi^{pqij}_{\alpha}\Phi^{rskl}_{\beta}\hat{\Omega}^{*}_{ijkl}g_{pqrs}^{(2,2)}
(\tilde{\lambda}^{\alpha}\lambda^{\beta})
+\mathrm{c.c.}+\cdots
~.
}
We see that generally, for $g\neq0$, all complex structure moduli obtain masses.

\subsection{Effective 3d $\ncal=1$ supergravity}\label{sec:eff3d}

From the previous analysis we have established the existence of the following massive $\ncal=1$ supermultiplets:
\boxedeq{
\mathrm{gravitino:}~\big(\hat{\chi}_{\mu};\mathcal{A}_{\mu}\big)~;~~~
\mathrm{vector:}
~\big(\mathcal{A}_\mu^{\prime A};\lambda^{\prime A+}\big)~;~~~
\mathrm{scalar:}~\big(\hat{M}^{\prime A};\lambda^{\prime A-}\big)~,~\big(Z^\alpha;\lambda^{\alpha}\big)~,\nn
}
while below the mass of the partial susy breaking we have the following massless $\mathcal{N}=1$ supermultiplets:
\boxedeq{
\mathrm{gravity:}~\big(g_{\mu\nu};{\chi}^{\prime +}_{\mu}\big)~;~~~
\mathrm{scalar:}~\big(V;\lambda^{-}\big)~,~\big(N^I;\lambda^{I}\big)~.
}
To lowest order in a scheme where the massive multiplets are integrated out\footnote{This simply corresponds to setting all massive multiplets to zero. Carrying out the integrating-out of the massive multiplets beyond leading order is outside the scope of the present paper, and is likely to be very difficult in practice. If one is to consider the  $\mathcal{N}=1$ theory as a Wilsonian effective action, the higher-dimensional couplings induced by integration over the high-momentum shells of the path integral cannot be determined unambiguously unless one is able to trace their origin in M-theory or in IIA string theory. However these next-to-leading order higher-derivative couplings are only partialy known.}, the theory describing the massless multiplets is an ungauged $\mathcal{N}=1$ supergravity. In order to explicitly obtain this theory by reduction from eleven dimensions we must include all terms with bosonic derivatives, in addition to the terms of sections \ref{sec:vacm}, \ref{sec:ft} which are obtained by considering the theory with all bosonic moduli frozen at their vacuum values.

There are two sources of such bosonic derivative terms: the first comes from the reduction of the $\psi^2G$ terms in eleven dimensions with $G\sim \d N^I\wedge\Psi^I+\mathrm{c.c.}$, cf. (\ref{g}),(\ref{cd}) and note that upon setting to zero the massive $Z^{\alpha}$ moduli the covariant derivative $D N^I$ reduces to an ordinary one. We thus obtain:
\eq{\spl{\label{m3}\frac{1}{2\kappa^2}&\int\d^{11}x\sqrt{-g}~\!\big[
-\frac{1}{48}
(\tilde{\psi}_M\Gamma^{MNPQRS}\psi_N+12\tilde{\psi}^P\Gamma^{QR}\psi^S)G_{PQRS}\big]\\
&\longrightarrow\frac{1}{2\kappa^2}\int\d^{3}x\sqrt{-g_{(3)}}~\hat{G}_{I\bar{J}}\big[
96(\tilde{\lambda}^{-}\gamma^{\mu}\lambda^I)
\partial_{\mu}\bar{N}^J
+16(\tilde{\lambda}^{I}\gamma^{\mu}\gamma^{\nu}{\chi}^{\prime +}_{\mu})
\partial_{\nu}\bar{N}^J\big]
+\mathrm{c.c.}~,
}}
where we took (\ref{ps}),(\ref{dde2}),(\ref{red}) into account.

The second source comes from the reduction of fermion kinetic terms in eleven dimensions, by taking into account the decomposition of the eleven-dimensional spinor derivative:
\eq{\label{covdev}\nabla_m\psi=\hat{\nabla}_m\psi+\frac{1}{16}\partial_{\mu}\ln V(\gamma^{\mu}\otimes\gamma_m\gamma_9)\psi~,}
where $\hat{\nabla}_m\psi$ is the covariant spinor derivative of the CY fourfold. Note that in section \ref{sec:ft} we made no distinction between $\hat{\nabla}_m\psi$ and $\nabla_m\psi$. Indeed we see explicitly from (\ref{covdev}) that the two coincide for $V\rightarrow \mathring{V}$. The equation above can be easily derived by decomposing the eleven-dimensional spin connection  into its three- and eight-dimensional components. Explicitly we have:
\eq{\omega^{(11)}_{mnp}=\omega^{(8)}_{mnp}~;~~~ \omega^{(11)}_{\mu\nu\rho}=\omega^{(3)}_{\mu\nu\rho}~;~~~
\omega^{(11)}_{mn\mu}=\frac18 \partial_{\mu}\ln Vg_{mn}~;~~~
\omega^{(11)}_{m\nu\rho}=0~.
}
In section \ref{sec:ft} we show that freezing the bosonic moduli to their vacuum values implies the vanishing of $\gamma^m{\nabla}_{[m}\psi_{n]}$ and ${\nabla}_m\psi_{\mu}$. This is no longer the case here. 
Specifically taking (\ref{covdev}),(\ref{ps}),(\ref{red}) into account while setting to zero all massive fermions we obtain the following terms:
\eq{\label{m1}\frac{1}{2\kappa^2}\int\d^{11}x\sqrt{-g} 
~2\tilde{\psi}_{M}\Gamma^{MmR}\nabla_m\psi_R
\longrightarrow \frac{1}{2\kappa^2}\int\d^{3}x\sqrt{-g_{(3)}} \left[9\partial_{\mu}V(\tilde{\lambda}^{-}\gamma^{\nu}\gamma^{\mu}{\chi}^{\prime +}_{\nu})\right]~,
}
in addition to the 3d kinetic fermion terms already derived in section \ref{sec:ft}:
\eq{\label{m2}
\frac{1}{2\kappa^2}\int\d^{3}x\sqrt{-g_{(3)}} ~V\Big\{(\tilde{\chi}^{\prime +}_{\mu}\gamma^{\mu\nu\rho}\nabla_{\nu}\chi^{\prime+}_{\rho})
+3^22^3(\tilde{\lambda}^{-}\gamma^{\nu}\nabla_{\nu}{\lambda}^{-})
+2^7\left[\hat{G}_{I\bar{J}}
(\tilde{\lambda}^{I}\gamma^{\nu}\nabla_{\nu}{\lambda}^{J c})
+\mathrm{c.c.}\right]
\Big\}
~.}
The latter are obtained from (\ref{j5}) by setting to zero the massive fermions, integrating over the CY fourfold and taking (\ref{dde2}) into account.

Up to quartic fermion terms, the complete fermionic action is given by the sum of (\ref{m3}), (\ref{m1}), (\ref{m2}). 
In order for the kinetic terms to be canonical we need  to rescale the fermions in addition to the Weyl rescaling (\ref{wr}). Specifically we require the following rescalings:
\eq{\label{fres}
g_{\mu\nu}\rightarrow V^{-2}g_{\mu\nu}~; ~~~{\chi}^{\prime +}_{\mu}\rightarrow V^{-\frac12}{\chi}^{\prime +}_{\mu}
~; ~~~\lambda^-\rightarrow 3^{-1}2^{-\frac32}V^{\frac12}\lambda^-~; ~~~\lambda^I\rightarrow 2^{-\frac72}V^{\frac12}\lambda^I
~,}
upon which the three-dimensional action becomes:
\eq{\spl{\label{m}
\frac{1}{2\kappa^2}&\int\d^{3}x\sqrt{-g_{(3)}} ~\!\Big\{
{R}-\frac98\partial_{\mu}\ln V\partial^{\mu}\ln V
-2\hat{G}_{I\bar{J}}\partial_{\mu}N^I\partial^{\mu}\bar{N}^J\\
+&(\tilde{\chi}^{\prime +}_{\mu}\gamma^{\mu\nu\rho}\nabla_{\nu}\chi^{\prime+}_{\rho})
+(\tilde{\lambda}^{-}\gamma^{\nu}\nabla_{\nu}{\lambda}^{-})
+2\hat{G}_{I\bar{J}}
(\tilde{\lambda}^{I}\gamma^{\nu}\nabla_{\nu}{\lambda}^{J c})
\\
+&\frac{3}{2\sqrt{2}}\partial_{\mu}\ln V(\tilde{\lambda}^{-}\gamma^{\nu}\gamma^{\mu}{\chi}^{\prime +}_{\nu})
+\big[\hat{G}_{I\bar{J}}(\tilde{\lambda}^{I}\gamma^{\mu}\gamma^{\nu}{\chi}^{\prime +}_{\mu})
\partial_{\nu}\bar{N}^J
+\frac{1}{\sqrt{2}}\hat{G}_{I\bar{J}}(\tilde{\lambda}^{-}\gamma^{\mu}\lambda^I)
\partial_{\mu}\bar{N}^J
+\mathrm{c.c.}\big]
\Big\}
~,}}
where  we have 
reinstated the kinetic terms for the massless bosonic moduli from (\ref{sb3d}) and we have also rescaled $N^I\rightarrow N^I/\sqrt{2}$ as before. In deriving the above form of the action one has to take into account the effect of the Weyl transformation on the spinorial derivative:
\eq{ 
\nabla_{\nu}\chi\rightarrow \nabla_{\nu}\chi-\frac12\gamma_{\nu\rho}\chi~\!\partial^{\rho}\!\ln V
~.}
Note also that below the scale of the partial supersymmetry breaking the volume $V$, the bosonic superpartner of $\lambda^-$, is the only massless bosonic modulus among the $M^A$'s. The fact that its kinetic term is not canonically normalized can be understood as follows: Parameterizing the K\"{a}hler moduli as a function of $V$ we have:
\eq{\label{e}
M^A=\left(\frac{V}{\mathring{V}}\right)^{\frac14}\mathring{M}^A
~;~~~
\mathring{M}^A\mathring{M}^BG_{AB}=2\bigg(\frac{\mathring{V}}{V}\bigg)^{\frac12}
~,
}
where we took (\ref{411}),(\ref{v}),(\ref{mc}) into account and 
we have reinstated $\mathring{V}$ which was hitherto set equal to one. Inserting the above in (\ref{sb3d}) we arrive at the kinetic term for $V$ appearing in (\ref{m}).

To read off the geometry of the (classical) moduli space let us define
\eq{\label{vr}U:=\frac{3}{2\sqrt{2}}\ln V~;~~~{H}_{I\bar{J}}:=e^{\frac{U}{\sqrt{2}}}
\hat{G}_{I\bar{J}}
~,
}
so that ${H}_{I\bar{J}}$ is a constant metric, as follows from (\ref{dde}),(\ref{e}) and the fact that $d_{AI\bar{J}}$ is independent of 
the K\"{a}hler moduli. 
Then the metric $\mathcal{G}$ of the moduli space reads:
\eq{\label{msm}\mathcal{G}_{UU}\delta U^2+\big[\mathcal{G}_{I\bar{J}}\delta N^I\delta \bar{N}^J+\mathrm{c.c.}\big]=\delta U^2+2e^{-\frac{U}{\sqrt{2}}}H_{I\bar{J}}\delta N^I\delta \bar{N}^J~,}
as can be seen from (\ref{m}),(\ref{vr}). This is the geometry of a flat complex $h^{2,1}$-torus, parameterized by $N^I$, fibered over the real line parameterized by $U$; it is locally isometric to hyperbolic space.

{}Furthermore one can show that (\ref{m}) is of the general form of a 
three-dimensional $\mathcal{N}=1$ ungauged supergravity action \cite{deWit:2003ja,deWit:2004yr}. To see this note that the non-vanishing 
Christoffel symbols associated with the moduli-space metric (\ref{msm}) are given by:
\eq{\label{ch}
\Gamma^U_{I\bar{J}}=\frac{1}{2\sqrt{2}}\hat{G}_{I\bar{J}}~;
~~~\Gamma^I_{UJ}=\Gamma^{\bar{I}}_{U\bar{J}}=-\frac{1}{2\sqrt{2}}\delta^I_J
~,}
where we used the fact that ${H}_{I\bar{J}}$ is constant. Up to quartic fermion terms, the action can thus be rewritten as
\eq{\spl{\label{f}
&S_{\mathrm{3d}}^{\ncal=1}=\frac{1}{2\kappa^2}\int\d^{3}x\sqrt{-g_{(3)}} ~\!\Big\{
{R}-\mathcal{G}_{UU}\partial_{\mu}U\partial^{\mu}U
-2\mathcal{G}_{I\bar{J}}\partial_{\mu}N^I\partial^{\mu}\bar{N}^J\\
+&(\tilde{\chi}^{\prime +}_{\mu}\gamma^{\mu\nu\rho}\nabla_{\nu}\chi^{\prime+}_{\rho})
+\mathcal{G}_{UU}(\tilde{\lambda}^{-}\gamma^{\nu}\nabla_{\nu}{\lambda}^{-})
+2\mathcal{G}_{I\bar{J}}
(\tilde{\lambda}^{I}\gamma^{\nu}\nabla_{\nu}{\lambda}^{J c})
+\mathcal{G}_{UU}(\tilde{\lambda}^{-}\gamma^{\nu}\gamma^{\mu}{\chi}^{\prime +}_{\nu})\partial_{\mu}U\\
+&\big[\mathcal{G}_{I\bar{J}}(\tilde{\lambda}^{I}\gamma^{\mu}\gamma^{\nu}{\chi}^{\prime +}_{\mu})
\partial_{\nu}\bar{N}^J
+\mathcal{G}_{UU}\Gamma^U_{I\bar{J}}(\tilde{\lambda}^{-}\gamma^{\mu}\lambda^I)
\partial_{\mu}\bar{N}^J
+\mathcal{G}_{I\bar{J}}\Gamma^{\bar{J}}_{U\bar{K}}(\tilde{\lambda}^{I}\gamma^{\mu}\lambda^-)
\partial_{\mu}{N}^{\bar{K}}
+\mathrm{c.c.}\big]
\Big\}
~,}}
which is precisely of the expected form of ungauged three-dimensional supergravity as given in \cite{deWit:2003ja,deWit:2004yr}.\footnote{To identify (\ref{f}) of the present paper with the action given in  eq.~(2.23) of \cite{deWit:2003ja} one must set $N=1$ therein and accordingly drop the $I$ index, which also implies that one must set $Q^{IJ}_i=0$ in eq.~(2.24) of \cite{deWit:2003ja}. Furthermore the real scalars $\phi^i$ there are identified with $(U,\mathrm{Re}N^I,\mathrm{Im}N^I)$ here; the Majorana fermions  $\chi^{i}$ there are identified with $(\lambda^-,\lambda^{I\pm})$ here; the gravitino $\psi_{\mu}$ there is identified with ${\chi}^{\prime +}_{\mu}$ here; the metric $g_{ij}$ there is identified with the moduli-space metric $\mathcal{G}$ here.}

\section{F-theory lift}\label{sec:ftheory}

For the F-theory interpretation of the $\ncal=1$ vacuum we will use the approach developped in \cite{Grimm:2010ks,Grimm:2011sk}. We will first review this 
approach following closely the conventions and notation of these references.\footnote{For this section, and this section only, the index $\alpha$ will be used to enumerate the vertical divisors of the elliptically fibered CY fourfold; it should not be confused with the index $\alpha$ elsewhere in the text, which is used to enumerate the complex structure deformations of the CY.}

In section \ref{sec:eff} we considered $\mathcal{N}=2$ M-theory compactifications on a generic CY fourfold $X$. Suppose that $X$ is elliptically fibered, i.e. there exists a holomorphic projection $\pi:X\rightarrow B$ of $X$ onto a K\"{a}hler threefold $B$ with generic fiber an elliptic curve. Let us call the two one-cycles of the two-torus the A- and the B-cycle, $S^1_A$, $S_B^1$ respectively.  
In the limit of small elliptic fiber, to be defined more precisely in the following, considering fiberwise the compactification of M-theory on $S^1_A$, we obtain a type IIA string theory on $B\times S^1_B$. Further T-dualizing along $S_B^1$ gives type IIB on $B\times S^1_{B_T}$, where  $S_{B_T}^1$ is the T-dual of  $S_B^1$. The IIB theory thus obtained has a varying axio-dilaton given by the complex structure modulus of the elliptic fiber of $X$; it also contains  D7-branes wrapping divisors $S\subset B$, with $S$ given by the set of points of $B$ over which the elliptic fiber degenerates (generally this means that a one-cycle shrinks to a point). Taking the volume of the elliptic fiber to zero decompactifies $S_{B_T}^1$ resulting in a four-dimensional $\mathcal{N}=1$ IIB string theory compactification on $B$ with D7 branes and varying axio-dilaton \cite{Vafa:1996xn}. Away from certain special limits \cite{Sen:1996vd,Sen:1997gv}, this theory 
is nonperturbative and is refered to as `F-theory on $X$'.

The type of degeneration of the elliptic fiber over the divisor $S$ determines the F-theory gauge group $G$. 
In the case where $G$ is non-abelian, $X$ itself is singular. 
To be able to obtain the effective action of M-theory on $X$ we will require that the singularities of $X$ have been resolved by blow-up, leading to a smooth CY fourfold $\hat{X}$. For ADE groups  this process is well understood and entails replacing the singularities of $X$ by exceptional divisors $D_i$, $i=1,\dots,\mathrm{rank}(G)$, which are $\mathbb{P}^1$ bundles over $S$. At generic points of $S$ the intersection form of the $D_i$'s is given by the 
Cartan matrix $C_{ij}$ of $G$.

A four-dimensional $\mathcal{N}=1$ supergravity with a given gauge group $G$ and a given number of chiral- and vector-multiplets, $n_c$,$n_v$ respectively,  is determined by 
a K\"{a}hler potential $K$, a holomorphic superpotential $W$ and a holomorphic gauge coupling function $\tau$. Determining the effective action of F-theory means specifying $(G,n_c,n_v,K,W,\tau)$ in terms of geometric data of the elliptic fibration of $X$. 
In the absence of a microscopic formulation of nonperturbative IIB string theory, one must proceed indirectly from M-theory on $\hat{X}$, following the dualities described previously. The three-dimensional $\mathcal{N}=2$ effective action of M-theory on $\hat{X}$, whose bosonic part is given in (\ref{sb3d2}), 
should then match the three-dimensional $\mathcal{N}=2$ low-energy effective action $S_{3}^{\mathrm{F}}$ of F-theory on ${X}$ further compactified on $S^1_{B_T}$. Obtaining  $S_{3}^{\mathrm{F}}$ would entail integrating out all KK and winding modes of F-theory to a scale below the lowest massive mode. This would be difficult to achieve in practice but can be explicitly performed for certain protected couplings of Chern-Simons type \cite{Cvetic:2012xn,Cvetic:2013uta}.

As a result of the resolution $X\rightarrow \hat{X}$ only a broken (Coulomb) phase of the  F-theory effective action 
$S_{3}^{\mathrm{F}}$ is directly accessible from M-theory on $\hat{X}$, and will in general include non-zero abelian gaugings and circle fluxes. 
The part of $S_3^{\mathrm{F}}$ describing the Kaluza-Klein zeromodes is obtained by circle reduction of a four-dimensional 
$\mathcal{N}=1$ supergravity action \cite{Wess:1992cp}. In the Coulomb phase the bosonic part of the latter reads:
\eq{\label{fa}S^{\mathrm{F}}=
\int\d x^4\sqrt{-g_{(4)}}\big\{R
-\frac14 C_{ij}\big[\mathrm{Re}\tau F^i_{\mu\nu} F^{j\mu\nu}
+\mathrm{Im}\tau F^i_{\mu\nu} F^j_{\rho\sigma}\varepsilon^{\mu\nu\rho\sigma}\big]
-2K^{\mathrm{F}}_{T_{\alpha}\bar{T}_{\bar{\beta}}}\nabla_{\mu} T_{\alpha}\nabla^{\mu} \overline{T}_{\bar{\beta}}
+V
\big\}
~,}
where $\alpha=1,\dots,n_c$, $i=1,\dots,\mathrm{rk}(G)$, and $C_{ij}$ is the Catan matrix of $G$. 

\vfill\break

The covariant derivatives of the scalars contain abelian gaugings parameterized by constant imaginary matrices $X_{i\alpha}$ and, upon reduction on $S^1_{B_T}$, the gaugings induced by circle fluxes $m_{\alpha}$ \cite{Grimm:2011sk}:\footnote{Note the slight abuse of notation in (\ref{cg}): 
the $A^i$'s are 4d vectors  whereas the graviphoton $A^0$ is a 3d vector. Moreover the $A^i$'s can have non-vanishing vevs along the $S^1_{B_T}$; the definition of the circle fluxes in  (\ref{cg}) ensures that $m_{\alpha}$ is invariant under large gauge transformations:
\eq{\label{lg}
A^i\rightarrow A^i- p^i\d\theta~;~~~
\mathrm{Im}T_{\alpha}\rightarrow \mathrm{Im}T_{\alpha}+\theta p^i X_{i\alpha}~;~~~p^i\in\mathbb{Z}
~,}
where $\theta$ is the coordinate of $S^1_{B_T}$. 
In addition the covariant derivative is invariant under local 3d gauge transformations:
\eq{\label{locg}
A^0\rightarrow A^0- \d\lambda~;~~~
\mathrm{Im}T_{\alpha}\rightarrow \mathrm{Im}T_{\alpha}+m_{\alpha}\lambda 
~,}
which can also be understood geometrically as diffeomorphism invariance under  $\theta\rightarrow\theta+\lambda(x^{\mu})$, where $x^{\mu}$ is a three-dimensional coordinate \cite{Grimm:2011sk}.}
\eq{\label{cg}
\nabla T_{\alpha}=\d T_{\alpha}+i m_{\alpha}A^0+X_{i\alpha}A^i
~;~~~
m_{\alpha}:=\int_{S^1_{B_T}}\langle\nabla\mathrm{Im}T_{\alpha}\rangle
~,}
where $A_{\mu}^0$ is the graviphoton of the circle reduction and the brackets denote the vev.

To proceed with the comparison with the M-theory effective action we must refine the decompositions (\ref{411}),(\ref{cexp}) on the basis of forms on the resolved elliptically-fibered $\hat{X}$. For that we take into account that $H^{1,1}(\hat{X})$ is generated by $e_A=(e_0,e_{\alpha},e_{i})$, where:
\begin{itemize}
\item  $e_0$ is the Poincaré dual of $B$
\item $e_{i}$, $i=1,\dots,\mathrm{rank}(G)$, are Poincaré dual to the exceptional divisors $D_i$
\item  $e_{\alpha}$, $\alpha=1,\dots,h^{1,1}(B)$, are Poincaré dual to the `vertical divisors' $D_{\alpha}$ of $\hat{X}$, i.e. $D_{\alpha}$ is of the form $\pi^{-1}(D^b_{\alpha})$ with $D^b_{\alpha}$ a divisor of $B$ ~,
\end{itemize}
and we have assumed for simplicity that there are no 
additional rational sections. 
In particular we have: 
\eq{h^{1,1}(\hat{X})=1+\mathrm{rank}(G)+h^{1,1}(B)~.}
Accordingly (\ref{411}),(\ref{cexp}) give rise to $h^{1,1}(\hat{X})$ vector multiplets whose 4d F-theory duals are identified as follows, focusing on the bosonic part of the multiplets \cite{Grimm:2010ks}:
\begin{itemize}
\item $(M^0,A^0)$ lifts to the metric components $(g_{33},g_{3\mu})$
\item $(M^i,A^i)$ lift to the the 4d abelian vectors $A^i$
\item $(M^{\alpha},A^{\alpha})$ dualize to 3d scalar multiplets and lift to 4d chiral multiplets $T_{\alpha}$ 
\end{itemize}
In the following we will make use of the intersection numbers:
\eq{\spl{\label{in}
&\int_{\hat{X}}e_{\alpha}\wedge 
e_{\beta}\wedge e_{\gamma}\wedge e_{\delta}=
\int_{\hat{X}} e_{i}\wedge e_{\alpha}\wedge 
e_{\beta}\wedge e_{\gamma}=
\int_{\hat{X}} e_{0}\wedge e_{i}\wedge e_{A}\wedge 
e_{B}=0\\
K_{ij\alpha\beta}:=&\int_{\hat{X}}e_{i}\wedge e_{j}\wedge e_{\alpha}\wedge 
e_{\beta}=-C^{\gamma}_{ij}K_{\alpha\beta\gamma}~;~~~
K_{\alpha\beta\gamma}:=\int_{\hat{X}}e_{0}\wedge e_{\alpha}\wedge 
e_{\beta}\wedge e_{\gamma}
~,}}
where the explicit form of the coefficients $C^{\gamma}_{ij}$ will not be necessary in the following. Let us also note that:
\eq{\label{fg}c_1(B)=k^{\alpha}e_{\alpha}\vert_B~;~~~e_0\wedge e_0=-\pi^*c_1(B)\wedge e_0~,}
for some coefficients $k^{\alpha}$.

{}Furthermore we need to refine the decomposition (\ref{ge}) of the  M-theory field-strength $G$ on the basis of four-forms on $\hat{X}$. As we will see in the following, in order to make contact with the 4d gaugings and circle fluxes that appear on the F-theory side, it suffices to consider the vertical  part $G_V$ of the partially supersymmetry-breaking vacuum (\ref{vfn1}). Recall that the fourth cohomology of $\hat{X}$ splits into a horizontal and a vertical part: 
\eq{H^4(\hat{X})=H^4_H(\hat{X})\oplus H^4_V(\hat{X})~,}
where $H^4_H(\hat{X})$ is spanned by the complex-structure deformations of $\Omega$, such that
\eq{H^4_H=H^{4,0}\oplus H^{3,1}\oplus H_H^{2,2}\oplus H^{1,3}\oplus H^{0,4}~,}
with $H_H^{2,2}$ a subset of the primitive-(2,2) cohomology of $\hat{X}$. On the other hand  $H^4_V(\hat{X})$ is generated by products of two elements of $H^{1,1}(\hat{X})$ and is necessarily of type (2,2). Explicitly, we may expand:
\eq{\label{gv}G_V=N^{\alpha}e_0\wedge e_{\alpha}+\tilde{N}_{\alpha}\wedge \tilde{e}^{\alpha}+f^{i\alpha}e_i\wedge e_{\alpha}+f^{ij}e_i\wedge e_j~,}
where the $\tilde{e}^{\alpha}$'s are pullbacks of forms on $B$ that are `dual' to the ${e}_{\alpha}$'s, in the sense that
\eq{\label{d}\int_B e_{\alpha}\wedge\tilde{e}^{\beta}=\delta_{\alpha}^{\beta}~.}
Note that for $\hat{X}$ a CY with full $SU(4)$ holonomy (and not a subgroup thereof) $H^4(B)$ is generated by wedge products of two  elements of $H^{1,1}(B)$. In particular this implies that all 4d Poincar\'e-violating four-fluxes, i.e. those with two or no legs along the elliptic fiber, are contained in the vertical part of the cohomology.

Moreover each $\tilde{e}^{\alpha}$ can be written as a linear combination 
of ${e}_{\alpha}\wedge{e}_{\beta}$ and vice-versa. 
Taking (\ref{in}),(\ref{d}) into account this implies in particular:
\eq{
{e}_{\alpha}\wedge{e}_{\beta}=K_{\alpha\beta\gamma}\tilde{e}^{\gamma}
~.}
In general not all of the ${e}_{\alpha}\wedge{e}_{\beta}$'s are linearly independent: in (\ref{gv}) it is understood that the $\tilde{N}_{\alpha}$'s  correspond to independent linear combinations of ${e}_{\alpha}\wedge{e}_{\beta}$'s.

Neglecting the dynamics of the $h^{2,1}$ scalars $N^I$ for simplicity, the part of the M-theory effective action (\ref{sb3d2}) describing the dynamics of the vectormultiplets can be given in terms of the embedding tensor $\Theta_{AB}$ of (\ref{td}), and a K\"{a}hler potential 
$K:=-3\ln V$, where $V$ is given in (\ref{v}). Taking (\ref{in}),(\ref{fg}) into account we find:
\eq{K=\ln \hat{M}^0+\ln\left[\Big(
\frac16 \hat{M}^{\alpha}\hat{M}^{\beta}\hat{M}^{\gamma}
-\frac{1}{4\hat{M}^0}\hat{M}^{\alpha}\hat{M}^{\beta}\hat{M}^{i}\hat{M}^jC^{\gamma}_{ij}
\Big)
K_{\alpha\beta\gamma}+\cdots\right]
~,
}
where the ellipses denote terms that should vanish in order for the result to agree, after dualization, with  the zero-mode part of the F-theory action $S^F_3$. 
References \cite{Grimm:2010ks,Grimm:2011sk} find that this can be achieved  provided we take the following limit:\footnote{The scalars ${M}^A$ of the present paper correspond to  $(v^0,v^{\alpha},v^i)$ of \cite{Grimm:2010ks,Grimm:2011sk} whereas the $\hat{M}^A$'s defined in (\ref{kred}) correspond to $(R,L^{\alpha},\xi^i)$ of those references.}
\eq{\label{flim}
\varepsilon:=V^{-2}\rightarrow0~;~~~
\hat{M}^0\sim \varepsilon^{\frac32}~;~~~
\hat{M}^i\sim \varepsilon^{2}~;~~~
\hat{M}^{\alpha}\sim \varepsilon^{0}
~,}
while keeping only up to and including terms of order $\varepsilon^{\frac52}$, in addition to shifting:
\eq{ (\mathcal{A}^{\alpha},\hat{M}^{\alpha})\rightarrow
(\mathcal{A}^{\alpha},\hat{M}^{\alpha})
+\frac12 k^{\alpha}(\mathcal{A}^0,\hat{M}^0)
~.}
Note that this is equivalent to shifting $e_0\rightarrow e_0+\frac12 k^{\alpha}e_{\alpha}$. 
Similarly the components of $\Theta_{AB}$ can be computed using  (\ref{in}),(\ref{fg}):
\begin{eqnarray} \label{tt}
\Theta_{00}&=&\tfrac{1}{16}\big(N^\alpha- f^{ij}C^\alpha_{ij}\big)k^\beta k^\gamma{K}_{\alpha\beta\gamma}\,;\nn\\
\Theta_{0\alpha}&=&\tfrac14 \tilde N_\alpha-\cfrac18\big(N^\beta+ f^{ij}C^\beta_{ij}\big)k^\gamma{K}_{\alpha\beta\gamma}\,;\nn\\
\Theta_{\alpha i}&=&\tfrac14 \big(f^{jk}{K}_{\alpha ijk} -f^{j\beta}C^\gamma_{ij} {K}_{\alpha\beta\gamma}\big)\,;\nn\\
\Theta_{0i}&=&-\tfrac{1}{2} k^\alpha\Theta_{\alpha i}\,;\nn\\
\Theta_{\alpha\beta}&=&\tfrac14\big(N^\gamma- f^{ij}C^\alpha_{ij}\big){K}_{\alpha\beta\gamma}\,;\nn\\
\Theta_{ij}&=&\tfrac14\big(f^{k\alpha}{K}_{\alpha kij}+ f^{kl}{K}_{ijkl}-\tilde N_\alpha C^\alpha_{ij}\big)\,.
\end{eqnarray} 
For those to agree with the F-theory gaugings in (\ref{cg}) we must have \cite{Grimm:2011sk}:\footnote{A sign difference between the expressions in (\ref{tt}),(\ref{tc}) and those of \cite{Grimm:2011sk} is due to the sign difference between our definition (\ref{td}) of the embedding tensor and (3.31) of that reference.}
\eq{\label{tc}\Theta_{0\alpha}=\frac12 m_{\alpha}~;~~~\Theta_{i\alpha}=-\frac{i}{2}X_{i\alpha}~;~~~
\Theta_{00}=
\Theta_{0 i}=
\Theta_{\alpha\beta}=
\Theta_{ij}=0
~.}
More generally however, it has been argued that taking into account loop corrections in $S^F_3$ the only constraints that need to be imposed for the 
M-theory vacuum to admit an F-theory interpretation without 4d abelian gaugings are \cite{Cvetic:2012xn,Cvetic:2013uta}: 
\eq{\label{lc1}
\Theta_{\alpha\beta}=
\Theta_{i\alpha}=0~.}
In addition demanding the vanishing of  4d Poincar\'e{}-violating circle fluxes requires imposing the constraint:
\eq{\label{lc2}
\Theta_{0\alpha}=0~.}

\subsection*{The F-theory limit of the $\mathcal{N}=1$ vacuum}

We would now like to examine the contribution of the partially susy-breaking 
part of the fourform (\ref{vfn1}), i.e. the part proportional to $g$, to the 
constraints (\ref{lc1}),(\ref{lc2}). For that, only the vertical part of the fourform need to be considered: $G_V:= g\mathring{J}\wedge\mathring{J}$. Using (\ref{in}),(\ref{fg}) we compute:
\eq{\spl{\label{gv1}G_V=gV^2\Big(
e_0\wedge e_{\alpha}[2\hat{M}^0 \hat{M}^{\alpha}&-(\hat{M}^0)^2 k^{\alpha}]+
\tilde{e}^{\gamma}\hat{M}^{\alpha}\hat{M}^{\beta}K_{\alpha\beta\gamma}\\
&+2e_\alpha\wedge e_i\hat{M}^{\alpha}\hat{M}^{i}
+e_i\wedge e_j\hat{M}^{i}\hat{M}^{j}
\Big)~,}}
where  we have set $\mathring{M}\rightarrow M$ for simplicity of notation; it should be kept in mind however that the right-hand side above should be evaluated at the vacuum. We impose that the components of $G_V$, when expanded on the basis of integral cohomology $e_A$, should be finite in the limit (\ref{flim}); in addition we impose that $G_V$ should be non-vanishing in that limit. From these two requirements it follows that $g$ must scale as $g=V^{-2}g_f$, where $g_f$ is finite. Inserting this in (\ref{gv1}) and taking the limit (\ref{flim}) we find:
\eq{G_V\rightarrow g_f\hat{M}^{\alpha}\hat{M}^{\beta}K_{\alpha\beta\gamma}
\tilde{e}^{\gamma}
~.}
This is of the form (\ref{gv}) provided we make the identifications:
\eq{
\tilde{N}_{\gamma}=g_f\hat{M}^{\alpha}\hat{M}^{\beta}K_{\alpha\beta\gamma}~;~~~
N^{\alpha}=f^{i\alpha}=f^{ij}=0
~.}
From (\ref{tt}) we then compute:
\eq{\label{tcomp}\Theta_{0\gamma}=\frac14 g_f\hat{M}^{\alpha}\hat{M}^{\beta}K_{\alpha\beta\gamma}
~;~~~
\Theta_{ij}=- C^{\gamma}_{ij}\Theta_{0\gamma}~;~~~
\Theta_{i\alpha}=
\Theta_{00}=
\Theta_{0 i}=
\Theta_{\alpha\beta}=
0
~.}
We stress that, as remarked below (\ref{tc}), the non-vanishing of the $\Theta_{ij}$ components above is 
incompatible with the classical (tree-level) F-theory fluxes and requires taking loop corrections into consideration, cf. the last row of table (4.9) of ref.~\cite{Cvetic:2013uta}.

Comparing with (\ref{lc1}) we conclude that the partially supersymmetry-breaking vacuum is consistent with an F-theory interpretation without 4d abelian gaugings. 
Moreover from (\ref{lc2}) we see that it is 4d Poincar\'e{}-violating in general since it contains non-vanishing circle fluxes, as can be read off from (\ref{tc}),(\ref{tcomp}):
\eq{m_{\gamma}=\frac12 g_f\hat{M}^{\alpha}\hat{M}^{\beta}K_{\alpha\beta\gamma}~.}
This comes as no surprise for compactifications on smooth, full $SU(4)$-holonomy CY fourfolds. Indeed in this case it can be seen that four-fluxes that respect 
4d Poincar\'e{} invariance must necessarily obey $G\wedge e_A=0$,\footnote{Strictly-speaking this equation is only true in cohomology. However, since $G$ is harmonic in our approximation, it follows that $G\wedge e_A$ is harmonic and hence it also vanishes pointwise.} see e.g. \cite{Denef:2008wq}. This would in its turn imply the vanishing of the partial supersymmetry-breaking parameter $g$, cf.~(\ref{vfn1}), and the restoration of $\ncal=2$ supersymmetry. Our analysis based on the F-theory effective action has allowed us in particular to refine this discussion to include smooth resolutions of singular CY's.

\section{Conclusions}\label{sec:con}

We have derived the  3d $\ncal=1$ effective action in the large-volume limit, up to quartic fermion terms, describing  M-theory CY compactifications with two real supercharges. 
This is a theory of massless 3d $\ncal=1$ supermultiplets obtained by 
decoupling all massive fields below the partial supersymmetry-breaking scale. 
The theory is expected to be corrected by order-$l^6_P$ terms, and it would be interesting to examine to which extent these can be constrained.

It would also be interesting to examine whether there exists an $\ncal=1$ 3d supergravity which also incorporates the light massive supermultiplets.\footnote{Of course the massive 3d $\ncal=1$ supermultiplets can be described within 3d $\ncal=2$ supergravity via the super-Higgs mechanism.} 
Since the latter are of the order of the partial supersymmetry-breaking scale, such a putative supergravity would have to include a massive gravitino supermultiplet. In d$>$3 dimensions this is not believed to be possible, however since the three-dimensional case is somewhat degenerate there is a chance that 
such a supergravity exists.

One of the reasons why three-dimensional M-theory vacua are interesting is their relation 
to F-theory. 
The starting point of most F-theory constructions are the 3d $\ncal=2$ M-theory vacua that fall within the class of \cite{Becker:1996gj}.\footnote{Refs.~\cite{Bonetti:2013fma,Bonetti:2013nka} are recent exceptions.} On the other hand, from the perspective of eleven-dimensional supergravity, there is a very rich `landscape' of solutions (although it is unlikely that they can all be promoted to genuine M-theory vacua) and there seem to be many more possibilities that have not yet been  considered. The present paper was mainly motivated by this observation. It would be interesting to extend the search for F-theory duals beyond the paradigm of CY fourfolds.

\section*{Acknowledgment} We are grateful to Thomas Grimm, Henning Samtleben 
 and Matthias Weissenbacher for valuable discussions.

\appendix

\section{Spinor and gamma matrix conventions}\label{app1}

For a spinor $\psi$ in any dimension we define:
\eq{\widetilde{\psi}:=\psi^{Tr}C~,}
where $C$ is the charge conjugation matrix. In Lorentzian signatures, we also define
\eq{\overline{\psi}:=\psi^{\dagger}\G_{0}~,}
where the Minkowski metric is mostly plus. 
In all dimensions the Gamma matrices are taken to obey
\eq{
(\G^M)^{\dagger}=\G_0\G^M\G_0~.
}
Antisymmetric products of
Gamma matrices are defined by
\eq{
\G^{(n)}_{M_1\dots M_n}:=\G_{[M_1}\dots\G_{M_n]}~.
}

\subsection*{Three Lorentzian dimensions}

The charge conjugation matrix in $1+2$ dimensions satisfies
\eq{
C^{Tr}=-C; ~~~~~~ (C\g^\mu)^{Tr}=C\g^\mu; ~~~~~~ C^*=-C^{-1}~.
}
The fundamental (two-dimensional) spinor representation is real. 
 We define:
\eq{
\zeta^c:=\g_0C^{-1}\zeta^*~.
}
The Hodge-dual of an antisymmetric product of
gamma matrices is given by
\eq{
\star\g_{(n)}={{ (-1)^{\frac{1}{2}n(n-1)}}}\g_{(3-n)}~.
\label{hodge3}
}
For two anticommuting spinors $\chi$, $\varphi$ we have:
\eq{\label{a}\tilde{\chi}\gamma_{\mu_1\dots\mu_k}\varphi
=(-)^{\frac12 k(k+1)}\tilde{\varphi}\gamma_{\mu_1\dots\mu_k}\chi~.}
We also note  the 
following useful properties:
\eq{\label{u1}
(\gamma_{\mu}\chi)^c=\gamma_{\mu}\chi^c~;~~~\widetilde{\gamma_{\mu}\chi}
=-\tilde{\chi}\gamma_{\mu}
~,}
and
\eq{\label{u2}(\widetilde{\chi}\gamma_{\mu_1\dots\mu_k}\varphi)^*
=\widetilde{\chi^c}\gamma_{\mu_1\dots\mu_k}\varphi^c~;~~~
(\widetilde{\chi^c}\gamma_{\mu_1\dots\mu_k}\varphi)^*
=\widetilde{\chi}\gamma_{\mu_1\dots\mu_k}\varphi^c
~.
}

\subsection*{Eight Euclidean dimensions}

The charge conjugation matrix in $8$ dimensions satisfies
\eq{
C^{Tr}=C; ~~~~~~ (C\g^\mu)^{Tr}=C\g^\mu; ~~~~~~ C^*=C^{-1}~.
}
The fundamental (eight-dimensional, chiral) spinor representation is real.
In this paper we work with a complexified chiral spinor $\eta$ (i.e. eight complex degrees of freedom). We define:
\eq{
\eta^c:= C^{-1}\eta^*~.
}
The chirality matrix is defined by
\eq{
\g_9:=\g_1\dots\g_8 ~.
}
The Hodge-dual of an antisymmetric product of
gamma matrices is given by
\eq{
\star\g_{(n)}\g_9=(-)^{\frac{1}{2}n(n+1)}\g_{(8-n)}~.
\label{hodge8}
}

\subsection*{Eleven Lorentzian dimensions}

The charge conjugation matrix in $1+10$ dimensions satisfies
\eq{
C^{Tr}=-C; ~~~~~~ (C\G^M)^{Tr}=C\G^M; ~~~~~~ C^{*}=-C^{-1}~.
}
The fundamental (32-dimensional) spinor representation 
is real, where we define the reality condition by
\eq{\label{r}
\overline{\e}=\widetilde{\e}~.
}
We decompose the eleven-dimensional Gamma matrices as
\eq{\label{gd}
\left\{ \begin{array}{ll}
\G^{\mu}=\g^\mu\otimes \g_9 &, ~~~~~\mu=0, 1, 2\\
\G^m=\bbone\otimes \g^{m-2} &, ~~~~~ m=3\dots 10
\end{array} \right.
~.}
It follows that
\eq{
C_{11}=C_3\otimes C_8\g_9~.
}
Given a complex spinor $\e$ the 
combination $\e+\e^c$ is real, in the sense of (\ref{r}), where
\eq{\e^c:=\G_0C^{-1}\e^*~.}
In the case where the eleven-dimensional spinor $\e$ is of factorized form, $\e=\zeta\otimes\theta$ with 
$\zeta$ and $\theta$ three- and eight-dimensional spinors respectively,  
the complex conjugate of the tensor product $\e^c$ is given by the tensor product of the complex conjugates:
\eq{\e^c=\zeta^c\otimes\theta^c~.}

\section{$SU(4)$ structures}\label{app:su4}

As we will now review a nowhere-vanishing complex, chiral, pure spinor $\eta$ of unit norm in eight euclidean dimensions defines an $SU(4)$ structure. In eight euclidean dimensions not every complex chiral spinor is pure: the property of purity is equivalent to the condition
\eq{\label{p1}
\we\eta=0~.}
Let $\eta_R$, $\eta_I$ be the real, imaginary part of $\eta$ respectively.
We will impose the normalization:
\eq{\label{b2}
\eta=\frac{1}{\sqrt{2}}(\eta_R+i\eta_I)~;~~~~~\we_R\eta_R=\we_I\eta_I=1~,}
so that $\tilde{\eta}^c\eta=1$, and (\ref{p1}) is equivalent to $\eta_R$, $\eta_I$ being orthogonal to each other: $\we_R\eta_I=\we_I\eta_R=0$.

Let us  define a real two-form $J$ and a complex self-dual four-form $\hat{\Omega}$ through the spinor bilinears
\eq{\spl{\label{b3}
iJ_{mn}&=\widetilde{\eta^c}\g_{mn}\eta\\
\hat{\Omega}_{mnpq}&=\we\g_{mnpq}\eta~.
}}
It can then be shown by Fierzing that these forms obey:
\eq{\spl{\label{b4}
J\wedge\hat{\Omega}&=0\\
\frac{1}{16}\hat{\Omega}\wedge\hat{\Omega}^*&=\frac{1}{4!}J^4 =\mathrm{vol}_8~,
}}
up to a choice of orientation,
and hence define an $SU(4)$ structure. 
The reduction of the structure group  can alternatively be seen from the fact that $Spin(6)\cong SU(4)$ is the stabilizer inside $Spin(8)$ of the pair of orthogonal Majorana-Weyl unit spinors $\eta_R$, $\eta_I$.

As follows from the vanishing of the first Chern class of the CY, 
any other globally-defined holomorphic top form $\Omega$ is related  
to $\hat{\Omega}$ by multiplication by a complex constant. We can always gauge-fix the phase of $\eta$ so that the holomorphic top forms 
$\hat{\Omega}$, ${\Omega}$ are related by:
\eq{\label{b5}\hat{\Omega}=\frac{4}{|\Omega|}\Omega~.}
Note that $|\Omega|^2$ does not depend on the coordinates of the CY.

Raising one index of $J$ with the metric defines an almost complex structure:
\eq{\label{b}
J_m{}^pJ_p{}^n=-\delta_m^n
~.}
Using the almost complex structure
we can define the projectors
\eq{\label{hp}
(\Pi^{\pm})_m{}^n\equiv\frac{1}{2}(\delta_{m}{}^{n}\mp i J_m{}^n)
~,}
with respect to which $\Omega$ is holomorphic
\eq{
(\Pi^{+})_m{}^i\Omega_{inpq}=\Omega_{mnpq}~; ~~~~~(\Pi^{-})_m{}^i\Omega_{inpq}=0 ~.
}

It will be useful to have the decomposition of all tensors in terms of 
$su(4)$ modules. Under an $so(8)\rightarrow su(4)$ decomposition the two-form, the three-form, the self-dual and the anti-self-dual four-form of $so(8)$ decompose respectively as:
\eq{\spl{
\bf{ 28}&\rightarrow \bf{ (6\oplus {6})\oplus(1\oplus 15) }\nn\\
\bf{ 56}&\rightarrow \bf{ (4\oplus \bar{{4}})\oplus(4\oplus 20)\oplus(\bar{4}\oplus \bar{20}) }\nn\\
\bf{35^+}&\rightarrow \bf{(1\oplus 1)\oplus (6\oplus {6})\oplus 20'\oplus  1}\nn\\
\bf{35^-}&\rightarrow \bf{(10\oplus \bar{10})\oplus 15}~.
}}
Following the conventions of \cite{Prins:2013wza}, we explicitly decompose the forms as follows:

{$\bullet$  Real two-form}
\eq{\label{dec2}F_{mn}=f^{(1,1)}_{2|mn}+f_2J_{mn}+\left(f^{(2,0)}_{2|mn}+\mathrm{c.c.}\right)~,}
where $f^{(1,1)}_{2|mn}\sim \bf{15}$ is a real traceless (1,1)-form, $f_2\sim \bf{1}$ is a real scalar, $f^{(2,0)}_{2|mn}\sim \bf{6}\oplus\bf{6}$ is a complex (2,0)-form. Note that given a complex (2,0)-form $\varphi^{(2,0)}$ transforming
in the reducible module $\bf{6}\oplus\bf{6}$, one may form irreducible representations thereof by imposing
a pseudoreality condition:
\eq{\label{pseudo}
\varphi^{(2,0)}_{mn}=\frac18~\! e^{i\theta}\hat{\O}_{mn}{}^{pq}\varphi^{(0,2)}_{pq}
~,}
where $\theta\in S^1$ is an arbitrary phase.

{$\bullet$  Real three-form}
\eq{\label{dec3}F_{mnp}=f^{(2,1)}_{3|mnp}+3f^{(1,0)}_{3|[m}J_{np]}
+\tilde{f}^{(1,0)}_{3|s}\Omega^{s*}{}_{mnp}
+\mathrm{c.c.}~,}
where $f^{(2,1)}_{3|mnp}\sim \bf{20}$ is a complex traceless (2,1)-form, ${f}^{(1,0)}_{3|m}, \tilde{f}^{(1,0)}_{3|m}\sim \bf{4}$ are complex (1,0)-forms.

{$\bullet$  Real self-dual four-form}
\eq{\label{dec4s}F^+_{mnpq}=
f^{(2,2)}_{4|mnpq}+6f_{4}J_{[mn}J_{pq]}
+\left(6f^{(2,0)}_{4|[mn}J_{pq]}+\tilde{f}_4\Omega_{mnps}
+\mathrm{c.c.}\right)~,}
where $f^{(2,2)}_{4|mnpq}\sim \bf{20'}$ is a real traceless (2,2)-form,  $f_4\sim \bf{1}$ is a real scalar,
$f^{(2,0)}_{4|mn}\sim \bf{6}+\bf{6}$ is a complex (2,0)-form,
$\tilde{f}_4\sim (\bf{1}\oplus\bf{1})$ is a complex scalar.

{$\bullet$  Real anti self-dual four-form}
\eq{\label{dec4a}F^-_{mnpq}=6f^{(1,1)}_{4|[mn}J_{pq]}
+\left(f^{(3,1)}_{4|mnpq}+\mathrm{c.c.}\right)~,}
where $f^{(1,1)}_{4|mn}\sim \bf{15}$ is a real traceless (1,1)-form,
$f^{(3,1)}_{4|mnpq}\sim \bf{\bar{10}}$ is a complex traceless (3,1)-form.
The following identity can easily be shown:
\eq{\label{c13}
\Omega^*_{[s|}{}^{mnp}f^{(3,1)}_{4|q]mnp}=0
~.}
We also list the following Hodge-duals:
\eq{\spl{\label{hd}\star 1&=\frac{1}{4!}J^4~;~~~
\star J=\frac{1}{3!}J^3\\
\star {f}^{(1,1)}&=-\frac{1}{2}{f}^{(1,1)}\wedge J^2
~;~~~\star {f}^{(2,0)}=\frac{1}{2}{f}^{(2,0)}\wedge J^2\\
\star {f}^{(1,2)}&=i{f}^{(1,2)}\wedge J~;~~~
\star \big({f}^{(1,0)}\wedge J\big)=\frac{i}{2}{f}^{(1,0)}\wedge J^2
~.}}
The following useful identities can be proved  by Fierzing \cite{tsim}:
\eq{\spl{
\frac{1}{4!\times 2^4}~&\hat{\Omega}_{rstu}\hat{\Omega}^{*rstu}=1\\
\frac{1}{6\times 2^4}~&\hat{\Omega}_{irst}\hat{\Omega}^{*mrst}
=(\Pi^+)_{i}{}^{m}\\
\frac{1}{4\times 2^4}~&\hat{\Omega}_{ijrs}\hat{\Omega}^{*mnrs}
=(\Pi^+)_{[i}{}^{m}(\Pi^+)_{j]}{}^{n}\\
\frac{1}{6\times 2^4}~&\hat{\Omega}_{ijkr}\hat{\Omega}^{*mnpr}
=(\Pi^+)_{[i}{}^{m}(\Pi^+)_{j}{}^{n}(\Pi^+)_{k]}{}^{p}\\
\frac{1}{4!\times 2^4}~&\hat{\Omega}_{ijkl}\hat{\Omega}^{*mnpq}
=(\Pi^+)_{[i}{}^{m}(\Pi^+)_{j}{}^{n}(\Pi^+)_{k}{}^{p}(\Pi^+)_{l]}{}^{q}~,
\label{bfive}
}}
Moreover, we have
\eq{\spl{
\widetilde{\eta^c}\eta=1; &~~~~~\we\eta=0\\
\widetilde{\eta^c}\g_{mn}\eta=iJ_{mn}; &~~~~~\we\g_{mn}\eta=0\\
 \widetilde{\eta^c}\g_{mnpq}\eta=-3J_{[mn}J_{pq]}; &~~~~~\we\g_{mnpq}\eta=\hat{\Omega}_{mnpq}\\
\widetilde{\eta^c}\g_{mnpqrs}\eta=-15iJ_{[mn}J_{pq}J_{rs]}; &~~~~~\we\g_{mnpqrs}\eta=0\\
\widetilde{\eta^c}\g_{mnpqrstu}\eta=105J_{[mn}J_{pq}J_{rs}J_{tu]}; &~~~~~\we\g_{mnpqrstu}\eta=0 ~,
\label{usefids}
}}
where we have made use of the identities
\eq{\spl{\label{jvol}
\sqrt{g}\; \varepsilon_{mnpqrstu}J^{rs}J^{tu}&=24J_{[mn}J_{pq]}\\
\sqrt{g} \;\varepsilon_{mnpqrstu}J^{tu}&=30J_{[mn}J_{pq}J_{rs]}\\
\sqrt{g} \;\varepsilon_{mnpqrstu}&=105J_{[mn}J_{pq}J_{rs}J_{tu]}
~.
}}
Note that the bilinears
$\we\g_{(p)}\eta$, ~$\widetilde{\eta^c}\g_{(p)}\eta$, vanish for $p$ odd.
The last line of equation (\ref{bfive}) together with the last line of the
equation above imply
\eq{\label{omvol}
\hat{\Omega}_{[ijkl}\hat{\Omega}^*_{mnpq]}=\frac{8}{35}\sqrt{g}\; \varepsilon_{ijklmnpq}~.
}
%
%
%
Finally we also list  the following relations:
\eq{\spl{
\g_m\eta&=(\Pi^+)_{m}{}^{n}\g_n\eta\\
\g_{mn}\eta&=iJ_{mn}\eta -\frac{1}{8}\hat{\Omega}_{mnpq}\g^{pq}\eta^c   \\
\g_{mnp}\eta&=3iJ_{[mn}\g_{p]}\eta
-\frac{1}{2}\hat{\Omega}_{mnpq}\g^q\eta^c\\
\g_{mnpq}\eta&=-3J_{[mn}J_{pq]}\eta -\frac{3i}{4}J_{[mn}\hat{\Omega}_{pq]ij}\g^{ij}\eta^c
+\hat{\Omega}_{mnpq}\eta^c
~.
\label{fierzsu}
}}
The action of $\gamma_{m_1\dots m_p}$, $p\geq 5$, on $\eta$ can be related to
the above formul{\ae}, using
the Hodge properties of gamma matrices given in appendix \ref{app1}.

With the help of (\ref{fierzsu}) and the tensor decompositions (\ref{dec4s}),(\ref{dec4a}), the following useful relations can be shown:
\eq{\spl{\label{c4}
\underline{F_4} \eta &= -12 f_4 \eta + 16\tilde{f}^{*}_{4} \eta^c - \frac{i}{8} f{^{(0,2)}_{4|mn}} \hat{\Omega}^{mnpq} \gamma_{pq} \eta^c\\
  \underline{F}_4 \g_m \eta &=
- 4i f_{4|mn}^{(1,1)}\g^n  \eta
+ \frac{1}{6}  f_{4|mnpq}^{(1,3)} \g_r \hat{\O}^{npqr} \eta^c~,
}}
where we define $\underline{A}:=\frac{1}{p!}A_{n_1\dots n_p}\gamma^{n_1\dots n_p}$ for any $p$-form $A$. 
In order to bring  the self-dual four-form vacuum solution (\ref{vfn1}) in the form of (\ref{dec4s}) one must set $f{^{(0,2)}_{4}}=f_{4}^{(1,1)}=f_{4}^{(1,3)}=0$ and $\tilde{f}_{4}=\frac34 f_4$ together with $f_4= g$, $f_4^{(2,2)}= g^{(2,2)}$. Specializing (\ref{c4}) to that case we thus obtain:
\eq{\label{c4s}
\underline{G}^{(1)}\eta = -12 g (\eta -\eta^c) ~;~~~
  \underline{G}^{(1)} \g_m \eta =0~.
}

\section{Expansion basis for the light modes}\label{sec:exp}

We will here give the details of the light-mode expansions used in the 
main text. As already mentioned in section \ref{sec:eff}, up to terms of order $l^6_P$, the light-modes coincide with the moduli of a fluxless compactification on a CY fourfold. In the next two subsections we will analyze separately the case of bosonic and fermionic modes.

\subsection{Bosonic light modes}\label{sec:expb}

In order to determine the bosonic moduli we must determine the variations of the bosonic fields that take a fluxless vacuum  to another vacuum, up to terms of  quadratic order in the variations. More precisely, we expand 
\eq{\label{kk1}
g_{MN}=\mathring{g}_{MN}+\delta g_{MN}~;~~~C_{MNP}=\mathring{C}_{MNP}+\delta C_{MNP}~;~~~\psi_{M}=\mathring{\psi}_{M}+\delta \psi_{M}~,
}
where $(\mathring{g}_{MN}, \mathring{C}_{MNP}, \mathring{\psi}_{M})$ is the vacuum of section \ref{sec:sol}, so that
\eq{\spl{\label{46}
\mathring{g}_{mn}&=\mathring{g}_{mn}(y)~;~~~\mathring{g}_{\mu\nu}=\eta_{\mu\nu}~;
~~~\mathring{g}_{\mu m}=0\\
4\partial_{[m}\mathring{C}_{npq]}&= \mathring{G}_{mnpq}(y)~;~~~\mathring{C}_{\mu\nu\rho}=
\mathring{C}_{\mu\nu p}=\mathring{C}_{\mu np}=0\\
\mathring{\psi}_{M}&=0
~.}}
The requirement that the `nearby' field configuration $({g}_{MN}, {C}_{MNP}, {\psi}_{M})$ solves the eleven-dimensional equations of motion to linear order in the variations constrains the form of the latter. 
More specifically it is well-known that in the case of fluxless CY compactification the linearized eleven-dimensional Einstein equations determine the variations of the metric as follows:\footnote{We mostly follow the notation of \cite{Haack:1999zv}; one difference is in our definition of the Hodge star operator:
\eq{\spl{\label{star11}
\star_{11}\omega_{p}\, =\, \frac{1}{p!(11-p)!}\sqrt{-g_{(11)}}\,\varepsilon_{M_1\ldots M_{11}}\omega^{M_{12-p}\ldots M_{11}} \d x^{M_1}\wedge \ldots\wedge \d x^{M_{11-p}}\nonumber~,
}}
with $\varepsilon_{01\ldots 10}=1$, and similarly in three Lorentzian and eight Euclidian dimensions.

The $A$ index in (\ref{star11}) 
may be raised/lowered using a Euclidean metric, $e^A=\delta^{AB}e_{B}$, etc.}
\eq{\label{47}i\delta g_{a\bar{b}}= \sum_{A=1}^{h^{1,1}}\delta M^Ae^A_{a\bar{b}}(y)~;~~~
\delta g_{\bar{a}\bar{b}}=\sum_{\alpha=1}^{h^{3,1}}\delta Z^{\alpha} b^{\alpha}_{\bar{a}\bar{b}}(y)
~;~~~\delta g_{\mu a}=0
~,}
where $\{e^A_{a\bar{b}}(y),~A=1,\dots,h^{1,1}\}$ is a basis of harmonic (1,1) forms on the vacuum CY, while $b^{\alpha}_{\bar{a}\bar{b}}$ is related to the  basis of harmonic (3,1) forms $\{\Phi^{\alpha}_{abc\bar{d}}(y),~\alpha=1,\dots,h^{3,1}\}$ on the CY via 
\eq{\label{bdef}
b^{\alpha}_{\bar{a}\bar{b}}:=\frac{1}{3|\Omega|^2}\Omega^{*cde}{}_{\bar{a}}
\Phi^{\alpha}_{cde\bar{b}}~;~~~
|\Omega|^2:=\frac{1}{4!}\Omega_{abcd}\Omega^{*abcd}
~,}
and we have found it convenient to introduce holomorphic/antiholomorphic  internal indices from the beginning of the latin alphabet: $a,b,\dots=1,\dots,4$~;~ $\bar{a},\bar{b},\dots=1,\dots,4$. Note that 
$b^{\alpha}_{\bar{a}\bar{b}}$ defined in (\ref{bdef}) is automatically symmetric in 
its two lower indices, cf., (\ref{c13}). Using (\ref{bfive}) the above relation can be 
`inverted'  to give:
\eq{\label{inv}\Omega_{abc}{}^{\bar{d}}b^{\alpha}_{\bar{d}\bar{e}}=2
\Phi^{\alpha}_{abc\bar{e}}~.}

We may think of the vacuum metric $\mathring{g}_{mn}$ as being defined at a 
point $(\mathring{M}^A,\mathring{Z}^{\alpha})$ in the space of moduli $\mathfrak{M}$, while 
the nearby metric ${g}_{mn}$ is defined at the  
point 
\eq{({M}^A,{Z}^{\alpha}):=(\mathring{M}^A+\delta M^A,\mathring{Z}^{\alpha}+\delta Z^{\alpha})\in\mathfrak{M}~.}
The variations $(\delta M^A,\delta Z^{\alpha})$ span the cotangent space of $\mathfrak{M}$ at 
the point $(\mathring{M}^A,\mathring{Z}^{\alpha})$. In fact $\mathfrak{M}$ has a 
structure of direct product (at least locally): 
\eq{\mathfrak{M}=\mathfrak{M}_k\times \mathfrak{M}_c~,} 
where 
$M^A\in\mathfrak{M}_k$ parametrize the moduli space of K\"{a}hler deformations 
and $Z^{\alpha}\in\mathfrak{M}_c$ parametrize the moduli space of complex structure deformations. 

We will denote by $(\mathring{J},\mathring{\Omega})$ the K\"{a}hler and holomorphic forms of the vacuum CY defined at 
the point  $(\mathring{M}^{A},\mathring{Z}^{\alpha})\in\mathfrak{M}$, while those of the nearby CY defined at 
the point  $({M}^{A},{Z}^{\alpha})\in\mathfrak{M}$ will be denoted by $(J,\Omega)$.
The K\"{a}hler form depends linearly  
on the K\"{a}hler moduli: 
\eq{\label{411}J= \sum_{A=1}^{h^{1,1}}M^Ae^A~.}
In contrast the dependence of the holomorphic form on the complex structure moduli is more complicated to define due to the variation of Hodge structures: as we move around in $\mathfrak{M}_c$ a form that is (4,0) at  $\mathring{Z}^{\alpha}\in\mathfrak{M}_c$ will generally develop $(4-p,p)$, $p\neq0$, components.

Similarly for the threeform we expand
\eq{\label{cexp}
C=\mathring{C}+\Big(\sum_{I=1}^{h^{2,1}} N^I \Psi_I+\mathrm{c.c.}\Big)+\sum_{A=1}^{h^{1,1}}  \mathcal{A}^A\wedge e^A
~,}
where $\{\Psi^I_{ab\bar{c}}(y),~I=1,\dots,h^{2,1}\}$ is a basis of harmonic (2,1) forms on the vacuum CY and $\mathcal{A}^A=\mathcal{A}^A_{\mu}\d x^{\mu}$ are three-dimensional one-forms; $\mathring{C}$ obeys $\d \mathring{C}=\mathring{G}$. 
Note that on a CY fourfold with full $SU(4)$ holonomy (and not a subgroup thereof) the $\Psi^I$'s are primitive. Indeed if $\Psi^I$ were not primitive $\star \left(\Psi^I\wedge J^2 \right)$ would be a nontrivial  harmonic (0,1) form -- in contradiction with the Hodge diamond of a CY fourfold with full $SU(4)$ holonomy \cite{yau}. Furthermore, the $\Psi^I$'s are assumed to depend holomorphically on the complex structure moduli $Z^{\alpha}$ \cite{Haack:1999zv}:
\eq{\label{hdep}
\partial_{\alpha}\Psi^I=\sigma_{\alpha IK}(Z,\bar{Z})\Psi^K+\tau_{\alpha I\bar{K}}(Z,\bar{Z})\bar{\Psi}^K
~;~~~
\partial_{\alpha}\bar{\Psi}^I=0
~,}
where for consistency we must have:
\eq{\label{con}
\bar{\partial}_{\bar{\beta}}\sigma_{\alpha IK}
=-\tau_{\alpha I\bar{L}}\bar{\tau}_{\bar{\beta}\bar{L}K}~;~~~
\bar{\partial}_{\bar{\beta}}\tau_{\alpha I\bar{K}}
=-\tau_{\alpha I\bar{L}}\bar{\sigma}_{\bar{\beta}\bar{L}\bar{K}}
~.}

As already mentioned, the bosonic moduli (\ref{47}),(\ref{cexp}) are not lifted at order in $l^3_P$. The reason is that their mass terms enter 
quadratically in the action (and the equations of motion) and are therefore of order $l^6_P$. 
To verify this directly, let us note first that the variation of the four-form $\delta G=\d\delta C$ induced by the variation of the moduli in (\ref{cexp}) vanishes, since $\delta C$ is harmonic. It follows that the linearization of the eleven-dimensional Einstein equations around the $\ncal=1$ vacuum solution of section \ref{sec:sol} receives contributions from the flux at order $l_P^6$ or higher so that the expressions in (\ref{47}) remain valid at lower orders. 
By the same reasoning the equation of motion for the four-form, linearized around the $\ncal=1$ vacuum solution, reduces to
\eq{l_P^3~\!\d\delta_{g}\!\star \mathring{G}=\mathcal{O}(l_P^6)~,}
where on the left-hand side we have denoted by $\delta_{g}\star$ the variation of the Hodge star induced by (\ref{47}). Dropping the higher-order terms in $l_P$ this equation can be rewritten as
\eq{\label{5}
\d\star v=0~;~~~v:=\frac{1}{4!}\mathring{G}_{mnp}{}^t\delta g_{st}\d x^m\wedge\d x^n\wedge\d x^p\wedge\d x^s
~,}
where the  Hodge star $\star$ and the contraction on the right-hand side above are taken  with respect to the vacuum CY metric.
Eq.~(\ref{5}) can then be seen 
to be automatically satisfied thanks to the harmonicity of $\delta g$ and $\mathring{G}$.

\subsection{Fermionic light modes}\label{sec:expf}

The fermionic moduli are the superpartners of the bosonic moduli of section \ref{sec:expb}; together they form the 3d $\mathcal{N}=2$ supergravity multiplets described explicitly in section \ref{sec:eff}.

Up to terms of order $l^6_P$, the fermionic moduli of the fluxless 
CY fourfold solution are given by the following expansions of the 
eleven-dimensional gravitino $\psi_M$:
\eq{\spl{\label{ps}
\psi_m&=\lambda^I\Psi^I_{mnp}\gamma^{np}\eta^c
+\lambda^{Ic}\bar{\Psi}^I_{mnp}\gamma^{np}\eta\\
&~~~+\lambda^{\alpha}\Phi^{\alpha}_{mpqr}\hat{\Omega}^*_n{}^{pqr}\gamma^n\eta
+\lambda^{\alpha c}\bar{\Phi}^{\alpha}_{mpqr}\hat{\Omega}_n{}^{pqr}\gamma^n\eta^c\\
&~~~+\lambda^Ae^A_{mn}\gamma^n\eta+\lambda^{Ac}e^A_{mn}\gamma^n\eta^c~;\\
\psi_{\mu}&=\chi_{\mu}\eta+\chi^c_{\mu}\eta^c
~,}}
where as in section \ref{sec:expb}: $I=1,\dots,h^{2,1}$, $\alpha=1,\dots,h^{3,1}$, $A=1,\dots,h^{1,1}$; $\lambda^I$, $\lambda^{\alpha}$, 
 $\lambda^A$, are complex 3d spinors (four real components) and  
$\lambda^c$ denotes the complex 
conjugate of $\lambda$, cf., appendix \ref{app1}; $\chi_{\mu}$ will be identified with a complex 3d gravitino; $\eta$ is the covariantly constant spinor of the vacuum CY.

\section{Reduction}\label{app:reduction}

In this section we give the technical details leading up to the effective actions (\ref{sb3d2}) and (\ref{f}). 
Our definition of the Hodge star implies:
\eq{\label{hs}\star\varphi\wedge\omega= \d^Dx\sqrt{\pm g_{(D)}}~\!\varphi\cdot\omega~,}
where $\varphi$, $\omega$ are $p$-forms in $D$-dimensional space, and the plus, minus sign on the right-hand side above is for Euclidean, Lorentzian signature respectively; we have defined
\eq{\varphi\cdot\omega:=\frac{1}{p!}\varphi_{m_1\dots m_p}\omega^{m_1\dots m_p}~.}
The volume $V$ of the internal CY fourfold is defined with respect to the 
CY metric $g_{mn}$:
\eq{\label{v}V:=\int \d^8y\sqrt{g_{(8)}}=\frac{1}{4!}\int J^4~.}
Moreover we define:
\eq{\label{d4}V_{A_1\dots A_n}:=\frac{1}{4!}\int e_{A_1}\wedge \dots\wedge e_{A_n}\wedge J^{4-n}}
It follows that
\eq{\label{vids}\partial _{A}V=4V_{A}~;~~~\partial _{A}\partial _{B}V=12V_{AB}~;~~~\partial _{A}\partial _{B}\ln V=12\frac {V_{AB}}{V}-16\frac {V_{A}V_{B}}{V^{2}}
~,}
where $\partial _{A}:=\partial/\partial M^A$. 
Taking (\ref{hd}),(\ref{hs}) into account we have:
\eq{\label{d5}\frac{1}{3!}e_A\wedge J^3
=e_A\wedge \star J=\d^8y\sqrt{g_{(8)}}~\!e_A\cdot J
~.}
On the other hand $e_A\wedge J^3$ is a harmonic top form hence it is equal to 
a constant times the volume element of the CY. From (\ref{d4}),(\ref{d5}) it thus follows that 
\eq{\label{d6}e_A\cdot J=\frac{4V_A}{V}~.}
Furthermore expanding $e_A$ as in (\ref{dec2}) taking (\ref{d6}) into account 
we obtain:
\eq{\label{he}\star e_A=\frac{2}{3}\frac{V_A}{V}J^3-\frac12 e_A\wedge J^2~.}
We define:
\eq{\spl{\label{c9}G_{AB}:=\frac{1}{2V}\int e_A\wedge\star e_B
&=-\frac{1}{2V}\int \d^8y\sqrt{g_{(8)}}~\!e_A^{a\bar{b}} e_B^{c\bar{d}}g_{a\bar{d}}g_{c\bar{b}}\\
&=-6\frac {V_{AB}}{V}+8\frac {V_{A}V_{B}}{V^{2}}=-\frac12 \partial_A\partial_B\ln V
~,}}
where in the first line we used (\ref{hs}), while in the second line we took (\ref{he}),(\ref{d4}),(\ref{vids}) into account. 
Moreover from (\ref{d6}),(\ref{c9}) it follows that:
\eq{\spl{\label{c10}\int \d^8y\sqrt{g_{(8)}}~\!e_A^{a\bar{b}} e_B^{c\bar{d}}g_{a\bar{b}}g_{c\bar{d}}
=-12V_{AB}-2VG_{AB}=-16\frac {V_{A}V_{B}}{V}
~,}}
where we used $ig_{a\bar{b}}=J_{a\bar{b}}$. From (\ref{vids}) we also obtain: 
\eq{\label{c11}
V_{AB}\partial_{\mu} M^A\partial^{\mu} M^B =\frac{1}{12}V\partial_{\mu}
\ln V\partial^{\mu}\ln V-\frac{1}{6}
VG_{AB}\partial_{\mu} M^A\partial^{\mu} M^B 
~.}
From the fact that $e_A\wedge \star e_B$ is a harmonic top form it follows that it is equal to a constant times the volume element of the CY. From (\ref{hs}) and the definition (\ref{c9}) we thus get:
\eq{\label{ma}
e_A\cdot e_B= 2G_{AB}
~.}
Inserting the above into (\ref{d6}) and taking (\ref{411}) into account we arrive at:
\eq{\label{mb}V_A=\frac{V}{2}G_{AB}M^B~.}
Moreover contracting (\ref{ma}) with the K\"{a}hler moduli and using $J\cdot J=4$, which follows from (\ref{b}), we obtain:
\eq{\label{mc}\frac12 M^AM^BG_{AB}=1~;~~~M^AV_A=V~,}
where in the second equation we took (\ref{mb}) into account. 
It is also useful to define the following matrix:
\eq{\label{rd}R_{A}{}^{B}:=\delta_A^B-\frac{V_AM^B}{V}~.}
Using (\ref{mc}) one can show that $R$ is a projector and 
 $M^A$, $V_A$ are zero-eigenvectors on the left, right respectively:
\eq{\label{rprop}R^2=R~;~~~M^AR_{A}{}^{B}=0~;~~~R_{A}{}^{B}V_B=0~.}

Similarly we define:\footnote{This definition agrees with eq.~(38) of \cite{Haack:1999zv}; the minus sign on the right-hand side of (\ref{gij}) accounts for the difference in our definition of the Hodge star.}
\eq{\label{gij}
G_{I\bar{J}}:=-\frac14 \int \Psi_I\wedge\star\bar{\Psi}_J
=\frac18 \int \d^8y\sqrt{g_{(8)}}~\!
\Psi_I^{ac\bar{e}} \bar{\Psi}_J^{\bar{b}\bar{d}f}g_{a\bar{b}}
g_{c\bar{d}}g_{\bar{e}f}~,}
where we used (\ref{hs}), 
and
\eq{\label{daij}
d_{AI\bar{J}}:= \int e_A\wedge\Psi_I\wedge\bar{\Psi}_J
=\frac14 \int \d^8y\sqrt{g_{(8)}}~\!
e_A^{a_1\bar{b}_1}\Psi_I^{a_2a_3\bar{b}_2} \bar{\Psi}_J^{\bar{b}_3\bar{b}_4a_4}\varepsilon_{a_1\dots a_4}
\varepsilon_{\bar{b}_1\dots \bar{b}_4}
~.}
Then (\ref{411}) implies:
\eq{\label{dde}d_{AI\bar{J}}=4i\partial_AG_{I\bar{J}}~;~~~
G_{I\bar{J}}=-\frac{i}{4}M^Ad_{AI\bar{J}}~.}
Taking (\ref{hdep}) into account we also obtain:
\eq{\label{appcon}
\partial_{\alpha}G_{I\bar{J}}=\sigma_{\alpha IK}G_{K\bar{J}}~;~~~
\partial_{\alpha}d_{AI\bar{J}}=\sigma_{\alpha IK}d_{AK\bar{J}}
~.}
The metric on the space of complex structure moduli is defined as follows:\footnote{Eq.~(\ref{c14}) corrects a sign typo in (36) of \cite{Haack:1999zv}.}
\eq{\label{c14}G_{\alpha\bar{\beta}}:=-\frac{\int\Phi^{\alpha}\wedge\bar{\Phi}^{\beta}}
{\int\Omega\wedge\Omega^*}
=\frac{1}{4V}\int\d^8y\sqrt{g_{(8)}}~\!
b^{\alpha}_{\bar{a}\bar{b}}\bar{b}^{\beta}_{cd}g^{\bar{a}c}g^{\bar{b}d}
~.}
To show the second equality first note that $\Omega\wedge\Omega^*$ is a harmonic top form hence it is proportional to a constant times the volume element. From this, (\ref{hs}) and the fact that $\Omega$ is self-dual it follows that
\eq{\label{t}\int\Omega\wedge\Omega^*=V|\Omega|^2~.}
Note that $|\Omega|^2$ is constant with respect to the internal coordinates of the CY. 
Next, using (\ref{inv}),(\ref{bfive}),(\ref{hs}) and the fact that $\Phi^{\alpha}$ is anti-selfdual, one can show that 
\eq{\int\Phi^{\alpha}\wedge\bar{\Phi}^{\beta}
=-\frac{|\Omega|^2}{4}\int\d^8y\sqrt{g_{(8)}}~\!
b^{\alpha}_{\bar{a}\bar{b}}\bar{b}^{\beta}_{cd}g^{\bar{a}c}g^{\bar{b}d}~,}
from which the desired result follows. The metric can also be defined 
in terms of a K\"{a}hler potential:
\eq{\label{kpot}G_{\alpha\bar{\beta}}=\partial_{\alpha}\bar{\partial}_{\beta}K~;~~~
K(Z,\bar{Z}):=-\ln\int\Omega\wedge\Omega^*~.}
As follows from Yau's theorem, a Ricci-flat metric on a CY fourfold is uniquely determined by specifying a complex
structure and a K\"{a}hler class. Moreover complex structure and K\"{a}hler deformations are independent of each other, which implies that at least locally the moduli space $\mathfrak{M}$ of Ricci flat metrics  has a direct-product structure, $\mathfrak{M}=\mathfrak{M}_c\times\mathfrak{M}_k$, where $\mathfrak{M}_c$ and $\mathfrak{M}_k$ are the complex structure
and K\"{a}hler moduli spaces repsectively, see e.g. \cite{hthes} and references therein. A choice of $\Omega$, $J$ thus specifies a point in $\mathfrak{M}_c$, $\mathfrak{M}_k$ respectively, and hence a point $\mathfrak{M}$. 
As we move in  $\mathfrak{M}_c$ the holomorphic top form $\Omega$ varies holomorphically with $Z^{\alpha}$. Moreover rescalings of the form $\Omega\rightarrow f(Z)\Omega$, which  depend holomorphically on $Z^{\alpha}$ but do not depend on the CY coordinates, do not change the complex structure of the CY. Thus $\Omega$ may be viewed as a section of a holomorphic line bundle over  $\mathfrak{M}_c$ \cite{Strominger:1990pd}. Motion in  $\mathfrak{M}_c$ 
is described in terms of the K\"{a}hler-covariant derivative:
\eq{
\mathcal{D}_{\alpha}\Omega=\Phi^{\alpha}~;~~~
\delta\Omega=\delta Z^{\alpha}\mathcal{D}_{\alpha}\Omega
~,}
where $\mathcal{D}_{\alpha}:=\partial_{\alpha}+\partial_{\alpha}K$. 
The above is consistent with the definition of the K\"{a}hler potential in (\ref{kpot}) as can be seen by wedging both sides of the 
covariant derivative with $\Omega^*$ and integrating over the CY; it is also consistent with (\ref{47}),(\ref{bdef}). This follows by expressing the variation of the metric due to motion in  $\mathfrak{M}_c$ in terms of the variation of the complex structure, $\delta g_{mn}= -iJ_{mp}\delta J_{n}{}^{p}$, and taking into account
 that $\delta J_{\bar{a}}{}^{b}\propto\delta {Z}^{\alpha}\varepsilon^{bb_1\dots b_3}{\Phi}^{\alpha}_{\bar{a}b_1\dots b_3}$.

\subsection{Bosonic terms}\label{sec:bt}

\subsection*{Ricci scalar}

Inserting the metric ansatz (\ref{kk1}),(\ref{46}),(\ref{47}) in the eleven-dimensional Riemann tensor,
\eq{R^M{}_{NRS}=\partial_R\Gamma^M{}_{SN}-\partial_S\Gamma^M{}_{RN}
+\Gamma^M{}_{RT}\Gamma^T{}_{SN}-\Gamma^M{}_{ST}\Gamma^T{}_{RN}
~,}
we obtain:
\eq{\spl{
g^{\mu\nu}R^{\rho}{}_{\mu\rho\nu}&=\hat{R}\\
g^{\mu\nu}R^{r}{}_{\mu r\nu}&=-\frac12 g^{mn}\hat{\nabla}^2\delta g_{mn}
+\frac14 g^{mn}g^{pq}\partial^{\nu}\delta g_{mp}\partial_{\nu}\delta g_{nq}\\
g^{mn}R^{r}{}_{mrn}&=-\frac14 g^{mn}g^{pq}\partial_{\nu}\delta g_{mn}
\partial^{\nu}\delta g_{pq}
+\frac14 g^{mn}g^{pq}\partial^{\nu}\delta g_{mp}\partial_{\nu}\delta g_{nq}\\
g^{mn}R^{\rho}{}_{m\rho n}&=-\frac12 g^{mn}\hat{\nabla}^2\delta g_{mn}
+\frac14 g^{mn}g^{pq}\partial^{\nu}\delta g_{mp}\partial_{\nu}\delta g_{nq}
~,}}
where $\hat{R}$ and $\hat{\nabla}$ denote the Riemann scalar and the 
Laplacian of the three-dimensional metric $g_{\mu\nu}$, and we have taken into account that the internal metric $g_{mn}=\mathring{g}_{mn}+\delta g_{mn}$ satisfies the CY condition 
of Ricci-flatness. We thus obtain 
for the eleven-dimensional Ricci-scalar:
\eq{\label{rs}R=\hat{R}-\frac14 g^{mn}g^{pq}\partial^{\nu}\delta g_{mp}\partial_{\nu}\delta g_{nq}
-\frac14 g^{mn}g^{pq}\partial_{\nu}\delta g_{mn}
\partial^{\nu}\delta g_{pq}-
\hat{\nabla}_{\nu}\big(g^{mn}\partial^{\nu}\delta g_{mn} \big)
~.}
We emphasize that, provided the internal metric is Ricci-flat, the above expression is exact in the sense that the variations $\delta g$ above do not need to be infinitesimal. On the other hand noticing that (\ref{rs}) is quadratic in $\delta g_{mn}$ we conclude that in order to determine the latter we only need solve $R_{mn}(\mathring{g}+\delta g)=0$ to linear order in the variations. This leads to the expansions (\ref{47}). 
Passing to complex coordinates and using the expansions (\ref{47}) this gives:
\eq{\spl{\label{il}R=\hat{R}+ g^{a\bar{b}}g^{c\bar{d}}\partial^{\nu}M^A 
\partial_{\nu}M^B e^A_{a\bar{b}}e^B_{c\bar{d}}
&+\frac12 g^{a\bar{d}}g^{c\bar{b}}\partial^{\nu}M^A 
\partial_{\nu}M^B e^A_{a\bar{b}}e^B_{c\bar{d}}\\
&-\frac12 g^{a\bar{b}}g^{c\bar{d}}\partial^{\nu}Z^{\alpha} 
\partial_{\nu}\bar{Z}^{\beta} {b}^{\alpha}_{\bar{b}\bar{d}}\bar{b}^{\beta}_{ac}-
\hat{\nabla}_{\nu}\big(g^{mn}\partial^{\nu}\delta g_{mn} \big)
~.}}
Integrating over the CY coordinates, taking into account the definitions (\ref{v}),(\ref{c9}),(\ref{c14}), and the identities (\ref{c10}),(\ref{c11}), we arive at:
\eq{\label{a1}\int\d^{11}x\sqrt{-g_{(11)}} ~R= \int\d^3x \sqrt{-g_{(3)}}\big(V\hat{R}-VG_{AB}\partial_{\nu}M^A\partial^{\nu}M^B
-2VG_{\alpha\bar{\beta}}
\partial_{\nu}Z^{\alpha}\partial^{\nu}\bar{Z}^{\beta}+V
\partial_{\nu}\ln{V}\partial^{\nu}\ln{V}\big)~,}
where we have partially integrated over the last term in (\ref{il}) using the identity 
\eq{\partial_{\rho}\sqrt{g}
=\frac{1}{2}\sqrt{g} g^{mn}\partial_{\rho} g_{mn}~.}
Next we perform a Weyl rescaling in order to bring the 3d action to a canonical form,
\eq{\label{wr}g=e^{2\sigma}g'\Longrightarrow e^{2\sigma}R(g)=R(g')-2(D-1)g^{\prime\mu\nu}\nabla_{\mu}\partial_{\nu}\sigma
-(D-1)(D-2)g^{\prime\mu\nu}\partial_{\mu}\sigma\partial_{\nu}\sigma
~,}
where $D$ is the dimension of spacetime. Setting $V=e^{-\sigma}$, (\ref{a1}) becomes
\eq{\label{a2}\int\d^{11}x\sqrt{-g_{(11)}} ~R= \int\d^3x \sqrt{-g'_{(3)}}\big(R(g')-g^{\prime\mu\nu}\big[ G_{AB}\partial_{\mu}M^A\partial_{\nu}M^B
+2G_{\alpha\bar{\beta}}
\partial_{\mu}Z^{\alpha}\partial_{\nu}\bar{Z}^{\beta}+
\partial_{\mu}\ln{V}\partial_{\nu}\ln{V}\big]\big)~.}

\subsection*{Gauge kinetic terms}

From (\ref{cexp}),(\ref{hdep}) it follows that
\eq{\label{g}G=\mathring{G} 
+F^A\wedge e^A+\big(DN^I\wedge\Psi^I+\mathrm{c.c.}\big)
~,}
where $F^{A}:=\d\mathcal{A}^A$ and 
\eq{\label{cd}DN^I:=\d N^I+\d Z^{\alpha}N^J\sigma_{\alpha J I}
+\d\bar{Z}^{\alpha}\bar{N}^J\bar{\tau}_{\bar{\alpha}\bar{J}I}~.}
Inserting this in (\ref{sb}), applying the same Weyl rescaling (\ref{wr}) as before, the reduction of the gauge kinetic term gives
\eq{\label{fred}
 -\frac12\int\d^{11}x\sqrt{-g_{(11)}} ~ G^2= \int\d^3x \sqrt{-g'_{(3)}}\big(
-\frac12 V^2 g^{\prime\mu\nu}g^{\prime\rho\sigma}G_{AB}F_{\mu\rho}^{A}F^{B}_{\nu\sigma}
-4V^{-1}g^{\prime\mu\nu}G_{I\bar{J}}D_{\mu}N^ID_{\nu}\bar{N}^J
\big)
~.}

\subsection*{Chern-Simons term}

Inserting (\ref{g}) into (\ref{sb}) and neglecting terms of order $\mathcal{O}(l_P^6)$, the reduction of the 
Chern-Simons term gives:
\eq{\label{csred}
-\frac16\int C\wedge G\wedge G= \int\big(
d_{AI\bar{J}}\mathcal{A}^A\wedge
 DN^I\wedge D\bar{N}^J
-2\Theta_{AB}\mathcal{A}^A\wedge F^B 
\big)
~,}
where we have defined following:
\eq{\label{td}T:=\frac{1}{4}\int \mathring{G}\wedge J^2
~;~~~\Theta_{AB}:=\frac12\partial_A\partial_BT=
\frac{1}{4}\int \mathring{G}\wedge e^A\wedge e^B
~.}

Putting together (\ref{a2}),(\ref{fred}),(\ref{csred}) and dropping the primes of the Weyl-rescaled terms to simplify the notation, the complete bosonic action reads,\footnote{This is in agreement with \cite{Haack:1999zv,Haack:2001jz} up to an apparent minus sign difference in the last term and a difference in the normalization of $N^I$; more specifically we have: $N^{\mathrm{here}}=N^{\mathrm{there}}/\sqrt{2}$. Note however that 
the {sign} of $\varepsilon^{01\dots 10}$ is not specified in these references. For example (A.1) of \cite{Haack:1999zv} is inconsistent with a Minkowski-signature metric. On the other hand taking into account that, 
$$
\int d_{AI\bar{J}}\mathcal{A}^A\wedge
DN^I\wedge D\bar{N}^J=
\frac12\int d_{AI\bar{J}}\left(
N^I D\bar{N}^J -\bar{N}^J DN^I \right)\wedge F^A~,
$$
the sign of the last term of (\ref{sb3d}) can be seen to be in agreement with eq.~(3.34) of \cite{Berg:2002es}. 
}
\eq{\spl{\label{sb3d}
S^b=\frac{1}{2\kappa^2}\int\d^3x\Big\{ \sqrt{-g_{(3)}}\big(
{R}&-G_{AB}\partial_{\nu}M^A\partial^{\nu}M^B
-2G_{\alpha\bar{\beta}}
\partial_{\nu}Z^{\alpha}\partial^{\nu}\bar{Z}^{\beta}-
\partial_{\nu}\ln{V}\partial^{\nu}\ln{V}\\
&-\frac12 V^2 G_{AB}F_{\mu\nu}^{A}F^{B\mu\nu}
-4V^{-1}G_{I\bar{J}}D_{\mu}N^ID^{\mu}\bar{N}^J\big)\\
&-\varepsilon^{\mu\nu\rho}d_{AI\bar{J}}\mathcal{A}^A_{\mu}
D_{\nu}N^ID_{\rho}\bar{N}^J
+\varepsilon^{\mu\nu\rho}\Theta_{AB}\mathcal{A}_{\mu}^A F^B_{\nu\rho} 
+\mathcal{O}(l_P^6)\Big\}
~,}}
where for the Chern-Simons terms  we have taken into account that $\varepsilon^{012}=-1$. To put the above into a more canonical form we 
perform the following coordinate transformation:
\eq{\label{kred}
M^A\rightarrow \hat{M}^A:=V^{-1}(M)M^A
~,}
and we define the the corresponding K\"{a}hler form and volume function:
\eq{\hat{J}:=\hat{M}^Ae_A~;~~~\hat{V}(\hat{M}):=\frac{1}{4!}\int \hat{J}^4=V(\hat{M})=V^{-3}(M)~.}
In particular the derivative with respect to the new variables $\hat{M}^A$ is related to the derivative with respect to $M^A$ via:
\eq{
\hat{\partial}_A=-\frac{4}{3}V_AM^B\partial_B+V\partial_A
~,} 
from which obtain the expression for  the redefined K\"{a}hler metric:
\eq{\hat{G}_{AB}=-\frac12\hat{\partial}_A\hat{\partial}_B\ln\hat{V}
=V^2G_{AB}
~.}
Moreover we set in analogy to (\ref{dde}):
\eq{\label{dde2}\hat{G}_{I\bar{J}}:=-\frac{i}{4}\hat{M}^Ad_{AI\bar{J}}
=V^{-1}{G}_{I\bar{J}}=\frac14(\Psi_I\cdot\bar{\Psi}_J)
~,}
where the last equality above follows from (\ref{hs}),(\ref{gij}) and the fact that 
$\Psi_I\wedge\star\bar{\Psi}_J$ is a harmonic top form and hence equal to a constant times the volume element of the CY. 
After the above redefinitions and 
a further rescaling $N^I\rightarrow N^I/\sqrt{2}$, the bosonic action reads as in (\ref{sb3d2}), which we also reproduce here:
\eq{\spl{
S^b=\frac{1}{2\kappa^2}\int\d^3x \Big\{\sqrt{-g_{(3)}}\Big[
{R}&-\hat{G}_{AB}(\partial_{\nu}\hat{M}^A\partial^{\nu}\hat{M}^B+\frac12  F_{\mu\nu}^{A}F^{B\mu\nu})
\\
&-2G_{\alpha\bar{\beta}}
\partial_{\nu}Z^{\alpha}\partial^{\nu}\bar{Z}^{\beta}
-2\hat{G}_{I\bar{J}}D_{\mu}N^ID^{\mu}\bar{N}^J\Big]\\
&-\frac12\varepsilon^{\mu\nu\rho}d_{AI\bar{J}}\mathcal{A}^A_{\mu}
D_{\nu}N^ID_{\rho}\bar{N}^J
+\varepsilon^{\mu\nu\rho}\Theta_{AB}\mathcal{A}_{\mu}^A F^B_{\nu\rho} 
+\mathcal{O}(l_P^6)
\Big\}
~. }}
Note that the two-derivative terms above are already quadratic in the variations, as follows from the remark below (\ref{pro}). Hence within 
the quadratic approximation we may promote the couplings $G_{AB}(\mathring{M}),\dots$, which are evaluated at the vacuum, to full moduli-dependent couplings $G_{AB}(\mathring{M})\rightarrow G_{AB}({M})$, etc.

\subsection{Fermionic terms at the $\mathcal{N}=1$ vacuum}
\label{sec:ft}

For the purposes of this section we will fix all bosonic moduli to their vacuum values: $\Phi\rightarrow\mathring{\Phi}$. We will however omit the circles above the bosonic fields to keep the notation simple.

Taking into account the fact that $\Psi^I$, $\Phi^{\alpha}$, $e^A$ are harmonic and $J$, $\Omega$ are covariantly constant, it follows that 
\eq{\label{i4}\gamma^{n}\nabla_{[m}\psi_{n]}=0~.} 
Let us illustrate this for the terms in (\ref{ps}) proportional to the  $\lambda^{\alpha}$ modes, for which the derivation is slightly lengthier. Neglecting all irrelevant numerical factors, one 
expands:
\eq{\spl{
\gamma^{n}\nabla_{[m}\psi_{n]}
&\sim \gamma^{n}\nabla_{[m}\Phi^{\alpha}_{n]ijk}\hat{\Omega}^{*ijkl}\gamma_l\eta\\
&\sim \nabla_{[m}\Phi^{\alpha}_{n]ijk}\hat{\Omega}^{*ijkn}\eta
+\nabla_{[m}\Phi^{\alpha}_{n]ijk}\Pi^{+ni}\gamma^{jk}\eta^c\\
&\sim \nabla_{m}(\Phi^{\alpha}\cdot\hat{\Omega})\eta
+\nabla^{i}\Phi^{\alpha}_{mijk}\gamma^{jk}\eta^c+
i\nabla_{n}(\Phi^{\alpha}_{mijk}J^{ni})\gamma^{jk}\eta^c
+i\nabla_{m}(\Phi^{\alpha}_{nijk}J^{ni})\gamma^{jk}\eta^c
~,
}}
where we used (\ref{bfive}), (\ref{fierzsu}) to pass from the first to the second line; to go from the second to the third we used the vanishing of $\d\Phi^{\alpha}$ (which follows from the fact that $\Phi^{\alpha}$ is harmonic, the covariant constancy of $J$, $\Omega$, and the definition of the holomorphic projector (\ref{hp}). Moreover each term in the last line vanishes: 
the first by virtue of the fact that $\Phi^{\alpha}$ is a (3,1)-form; the second by virtue of the vanishing of $d^{\dagger}\Phi^{\alpha}$ which follows from the fact that $\Phi^{\alpha}$ is harmonic; the third term also vanishes for the same reason since it is equal to $\pm d^{\dagger}\Phi^{\alpha}$ depending on whether the index $m$ is holomorphic or antiholomorphic; the last term vanishes by virtue of the fact that $\Phi^{\alpha}$ is primitive. 
Similar manipulations  can be used to show (\ref{i4}) also for the terms in (\ref{ps}) proportional to the $\lambda^I$, $\lambda^{A}$ modes.

Taking (\ref{i4}) and the gamma-matrix decompostion (\ref{gd}) into account, the fermion kinetic terms in (\ref{sf}) reduce to:
\eq{\label{fk}
2\tilde{\psi}_{\mu}(\gamma^{\mu\nu\rho}\otimes\gamma_9)\nabla_{\nu}\psi_{\rho}+
4\tilde{\psi}_{m}(\gamma^{\nu\rho}\otimes\gamma^m)\nabla_{\nu}\psi_{\rho}
-2\tilde{\psi}_{m}(\gamma^{\nu}\otimes\gamma_9\gamma^{mp})\nabla_{\nu}\psi_{p}
~,}
where we also used that $\nabla_{m}\psi_{\mu}=0$, as follows from the spinor ansatz (\ref{ps}) and the fact that $\eta$ is covariantly constant.  
Moreover the first term in (\ref{fk}) gives:
\eq{\label{j1}2\tilde{\psi}_{\mu}(\gamma^{\mu\nu\rho}\otimes\gamma_9)\nabla_{\nu}\psi_{\rho}
=2(\tilde{\chi}^c_{\mu}\gamma^{\mu\nu\rho}\nabla_{\nu}\chi_{\rho})
+\mathrm{c.c.}
~.}
It is straightforward to see that the second term in (\ref{fk}) does not depend on the $\lambda^{\alpha}$, $\lambda^{I}$ modes: this is a consequence of the fact that the part of the spinor bilinear $\psi_m\otimes\eta$ that is linear in 
 $\Phi^{\alpha}$, $\Psi^{I}$ does not contain an $su(4)$ singlet. The same is true for the terms linear in the primitive part of $e^A$. More specifically we have:
\eq{
\label{j2}
4\tilde{\psi}_{m}(\gamma^{\nu\rho}\otimes\gamma^m)\nabla_{\nu}\psi_{\rho}
=8i(e^A\cdot J)
(\tilde{\lambda}^{Ac}\gamma^{\nu\rho}\nabla_{\nu}\chi_{\rho})
+\mathrm{c.c.}
~,}
where we used (\ref{ps}) and (\ref{fierzsu}). However the canonical form of fermion kinetic terms \cite{deWit:2003ja,deWit:2004yr} does not contain cross terms between $\lambda$ and $\chi_{\mu}$. This can be accomplished by redefining the gravitino as follows: 
\eq{\label{red}\chi_{\mu}=\chi'_{\mu}-2i(e^A\cdot J)\gamma_{\mu}\lambda^A~,}
so that the fermion kinetic terms do not contain cross terms between $\lambda$ and $\chi_{\mu}'$. 
Proceeding in a similar manner, the third term in (\ref{fk}) gives:
\eq{\spl{
\label{j3}
-2\tilde{\psi}_{m}(\gamma^{\nu}\otimes\gamma_9\gamma^{mp})\nabla_{\nu}\psi_{p}
&=\left[4(e^A\cdot e^B)-8(e^A\cdot J)(e^B\cdot J)\right]
(\tilde{\lambda}^{Ac}\gamma^{\nu}\nabla_{\nu}{\lambda}^{B})\\
&+3^22^8(\bar{\Phi}^{\bar{\alpha}}\cdot\Phi^{\beta})
(\tilde{\lambda}^{\alpha c}\gamma^{\nu}\nabla_{\nu}{\lambda}^{\beta})
+32(\bar{\Psi}^{\bar{I}}\cdot\Psi^{J})
(\tilde{\lambda}^{I c}\gamma^{\nu}\nabla_{\nu}{\lambda}^{J})
+\mathrm{c.c.}
}}
Putting together all the above we obtain the following kinetic fermion terms,
\eq{\spl{\label{j4}
2(\tilde{\chi}^{\prime c}_{\mu}\gamma^{\mu\nu\rho}\nabla_{\nu}\chi'_{\rho})&+\left[4(e^A\cdot e^B)+8(e^A\cdot J)(e^B\cdot J)\right]
(\tilde{\lambda}^{Ac}\gamma^{\nu}\nabla_{\nu}{\lambda}^{B})\\
&+3^22^8(\bar{\Phi}^{\bar{\alpha}}\cdot\Phi^{\beta})
(\tilde{\lambda}^{\alpha c}\gamma^{\nu}\nabla_{\nu}{\lambda}^{\beta})
+32(\bar{\Psi}^{\bar{I}}\cdot\Psi^{J})
(\tilde{\lambda}^{I c}\gamma^{\nu}\nabla_{\nu}{\lambda}^{J})
+\mathrm{c.c.}
~,}}
where we used the redefined gravitino, cf. (\ref{red}). 

In terms of the redefined fields (\ref{lr1}),(\ref{lr2}) the kinetic terms (\ref{j4}) can be written as:
\eq{\spl{\label{j5}
&2(\tilde{\chi}^{\prime c}_{\mu}\gamma^{\mu\nu\rho}\nabla_{\nu}\chi'_{\rho})+4(e^{\prime A}\cdot e^{\prime B})
(\tilde{\lambda}^{\prime Ac}\gamma^{\nu}\nabla_{\nu}{\lambda}^{\prime B})\\
+&3^22^8(\bar{\Phi}^{\bar{\alpha}}\cdot\Phi^{\beta})
(\tilde{\lambda}^{\alpha c}\gamma^{\nu}\nabla_{\nu}{\lambda}^{\beta})
+32(\bar{\Psi}^{\bar{I}}\cdot\Psi^{J})
(\tilde{\lambda}^{I c}\gamma^{\nu}\nabla_{\nu}{\lambda}^{J})
+\mathrm{c.c.}\\
+&3^22^3\left[(\tilde{\lambda}^{+}\gamma^{\nu}\nabla_{\nu}{\lambda}^{+})
+(\tilde{\lambda}^{-}\gamma^{\nu}\nabla_{\nu}{\lambda}^{-})\right]
~,}}
where we defined:
\eq{\label{dpm}\theta^{+}:=\theta+\theta^c~;~~~
\theta^{-}:=-i(\theta-\theta^c)
~,}
for any fermion $\theta$.

Inserting the decomposition (\ref{gd}) of the eleven-dimensional gamma matrices in the fermionic Lagrangian (\ref{sf}) we obtain the following mass terms,
\eq{\label{d34}
-\frac{1}{48}\mathring{G}_{pqrs}\Big[
\tilde{\psi}_{\mu}(\gamma^{\mu\nu}\otimes\gamma^{pqrs})\psi_{\nu}
+2\tilde{\psi}_{\mu}(\gamma^{\mu}\otimes \gamma_9\gamma^{npqrs})\psi_n
+24\tilde{\psi}^p(1\otimes\gamma^{qr}P_-)\psi^s
\Big]
~,}
where we used eight-dimensional Hodge duality (\ref{hodge8}) and taken into account the self-duality of the four-form flux at the vacuum; $P_\pm:=\frac{1}{2}(1\pm\gamma_9)$ are the eight-dimensional chirality projection operators. We must then insert in the above the decomposition (\ref{ps}) of the eleven-dimensional gravitino. Let us first list the following useful intermediate results:
\eq{\label{i1}
\mathring{G}_{pqrs}(\gamma^{\mu\nu}\otimes\gamma^{pqrs})\psi_{\nu}
=-3^2 2^5 g
\gamma^{\mu\nu}\chi_{\nu}\otimes(\eta-\eta^c)+\mathrm{c.c.}
~,}
where we took (\ref{vfn1}),(\ref{c4s}) into account. Moreover, 
\eq{\label{i2}
\mathring{G}_{pqrs}(\gamma^{\mu}\otimes \gamma_9\gamma^{npqrs})\psi_n
=-3^2 2^5 i g(e_A\cdot J)
\gamma^{\mu}\lambda^A\otimes(\eta-\eta^c)+\mathrm{c.c.}
~,}
where we used (\ref{bfive}), (\ref{fierzsu}). The fact that there are no terms on the right hand side other than the `trace part' of $e_A$ can also be understood by representation-theoretic arguments. For example, in order for the primitive part ${e}_A'$ of $e_A$ (which transforms in the ${\bf 15}$ of $su(4)$) to appear, the tensor product of $\mathring{G}\sim {\bf 20'\oplus 1}$ and  ${e}_A'\sim {\bf 15}$ should contain  either a singlet or an  (anti)holomorphic two-form (which transforms in the ${\bf 6}$ of $su(4)$). This is because the part of $\gamma^{npqrs}\psi_n$ proportional to $e_A$ contains an even number of gamma matrices, cf., (\ref{ps}), and thus can be brought using (\ref{fierzsu}) to the form $e_A\eta$, $e_A\eta^c$ or $e_A\gamma_{mn}\eta^c$. However  neither ${\bf 1}$ nor ${\bf 6}$ is in ${\bf (20'\oplus 1)\otimes 15}$. 
By an entirely analogous reasoning it can be seen that all terms proportional to $\Phi^{\alpha}\sim{\bf 10}$ vanish. 
Finally, a similar argument can be used to show that there can be no terms containing $\Psi_I$. In this case the part of $\gamma^{npqrs}\psi_n$ proportional to $\Psi_I$ contains an odd number of gamma matrices, cf., (\ref{ps}), and thus can be brought using (\ref{fierzsu}) to the form $\Psi_I\gamma_m\eta$ or $\Psi_I\gamma_m\eta^c$. Hence for $\Psi^I$ to appear, the tensor product of $\mathring{G}\sim {\bf 20'\oplus 1}$ and  $\Psi_I\sim {\bf 20}$ should contain  either a holomorphic (${\bf 4}$) or an  antiholomorphic one-form (${\bf \bar{4}}$) -- which is not the case.  

The last term in (\ref{d34}) gives:
\eq{\spl{\label{i3}
\mathring{G}_{pqrs}\tilde{\psi}^p(1\otimes\gamma^{qr}P_-)\psi^s
=&2\left[e_A^{mn}e_B^{pq}~\!g_{mnpq}^{(2,2)}
+4g(e_A\cdot e_B)+8g(e_A\cdot J)(e_B\cdot J)
\right]
(\tilde{\lambda}^{Ac}\lambda^B)\\
&-24g\left[(e_A\cdot e_B)-(e_A\cdot J)(e_B\cdot J)
\right]
(\tilde{\lambda}^{A}\lambda^B)\\
&-36\Phi_{pqij}^{\alpha}\Phi_{rskl}^{\beta}\hat{\Omega}^{*ijkl}g^{pqrs}_{(2,2)}
(\tilde{\lambda}^{\alpha}\lambda^{\beta})
-3^32^9g\bar{\Phi}^{\bar{\alpha}}\cdot\Phi^{\beta}
(\tilde{\lambda}^{\alpha c}\lambda^{\beta})
+\mathrm{c.c.}
~.}}
As before, the fact that the $\Psi_I$ terms drop out can be seen by purely representation-theoretic arguments. First one notes that the $P_-$ projector on the left-hand side of (\ref{i3}) projects out all terms quadratic in $\Psi_I$; then one notes that there are no singlets in the decomposition of the tensor products $\Psi_I\otimes \Phi^{\alpha}\otimes \mathring{G}$ or $\Psi_I\otimes e^{A}\otimes \mathring{G}$. By a similar argument one shows that there can be no terms of the form $\Phi^{\alpha}\otimes\bar{\Phi}^{\bar{\beta}}\otimes g^{(2,2)}$. In the derivation of (\ref{i3}) we have made repeated use of (\ref{bfive}),(\ref{usefids}),(\ref{fierzsu}) as well as the following identities:
\eq{
A_{p[m}J^p{}_{n]}=0
~,}
where $A_{mn}$ is any (1,1)-form, 
and
\eq{\spl{\label{rz}\Phi_{\bar{a}bce}^{\alpha}\Phi_{\bar{d}fgh}^{\beta}
\hat{\Omega}^{*efgh}g^{\bar{a}bc\bar{d}}_{(2,2)}
&=\frac{3}{2}\Phi_{\bar{a}bef}^{\alpha}\Phi_{\bar{c}dgh}^{\beta}
\hat{\Omega}^{*efgh}g^{\bar{a}b\bar{c}d}_{(2,2)}\\
&=\frac{3}{8}\Phi_{pqij}^{\alpha}\Phi_{rskl}^{\beta}
\hat{\Omega}^{*ijkl}g^{pqrs}_{(2,2)}\\
&=\frac{1}{3}\Phi_{pqri}^{\alpha}\Phi_{sjkl}^{\beta}
\hat{\Omega}^{*ijkl}g^{pqrs}_{(2,2)}
~,}}
where the first line is expressed in terms of holomorphic/antiholomorphic indices whereas real indices are used in the third and fourth lines. Eq.(\ref{rz}) can be shown by using the fact that the full antisymmetrization of five holomorphic (or antiholomorphic) indices vanishes. In particular the second line above makes manifest that this expression is symmetric in $(\alpha,\beta)$, which  can also be seen from the fact that there is no singlet in the decomposition of  $\wedge^2{\bf 10\otimes 10'}$. Furthermore this symmetry property is indeed consistent with the symmetry of $(\tilde{\lambda}^{\alpha}\lambda^{\beta})$, cf., (\ref{i3}), as follows from (\ref{a}).

Putting together the above, we obtain the following mass terms:
\eq{\spl{\label{i5}
-&\left[e_A^{mn}e_B^{pq}~\!g_{mnpq}^{(2,2)}
+4g(e_A\cdot e_B)+80g(e_A\cdot J)(e_B\cdot J)
\right]
(\tilde{\lambda}^{Ac}\lambda^B)\\
&+g\left[12(e_A\cdot e_B)-84(e_A\cdot J)(e_B\cdot J)
\right]
(\tilde{\lambda}^{A}\lambda^B)\\
&+18\Phi_{pqij}^{\alpha}\Phi_{rskl}^{\beta}\hat{\Omega}^{*ijkl}g^{pqrs}_{(2,2)}
(\tilde{\lambda}^{\alpha}\lambda^{\beta})
+3^32^8g\bar{\Phi}^{\bar{\alpha}}\cdot\Phi^{\beta}
(\tilde{\lambda}^{\alpha c}\lambda^{\beta})
+\mathrm{c.c.}\\
&+6g(\tilde{\chi}_{\mu}^{\prime-}\gamma^{\mu\nu}\chi_{\nu}^{\prime-})
+36g(e_A\cdot J)(\tilde{\lambda}^{A+}\gamma^{\nu}\chi_{\nu}^{\prime-})
~,}}
where we took (\ref{dpm}) into account and we used the redefined gravitino (\ref{red}). 
In terms of the redefined fields (\ref{lr1}),(\ref{lr2}) the mass terms (\ref{i5}) can be written as:
\eq{\spl{\label{i6}
&-\left[e_{A}^{\prime mn}e_{B}^{\prime pq}~\!g_{mnpq}^{(2,2)}
+4g(e_{A}'\cdot e_{B}')
\right]
(\tilde{\lambda}^{\prime Ac}\lambda^{\prime B})
+12g(e_{A}'\cdot e_{B}')
(\tilde{\lambda}^{\prime A}\lambda^{\prime B})
\\
&+18\Phi_{pqij}^{\alpha}\Phi_{rskl}^{\beta}\hat{\Omega}^{*ijkl}g^{pqrs}_{(2,2)}
(\tilde{\lambda}^{\alpha}\lambda^{\beta})
+3^32^8g\bar{\Phi}^{\bar{\alpha}}\cdot\Phi^{\beta}
(\tilde{\lambda}^{\alpha c}\lambda^{\beta})
+\mathrm{c.c.}\\
&+6g(\tilde{\chi}_{\mu}^{\prime-}\gamma^{\mu\nu}\chi_{\nu}^{\prime-})
+2^43^2g(\tilde{\lambda}^{+}\gamma^{\nu}\chi_{\nu}^{\prime-})
-2^43^4g
(\tilde{\lambda}^{+}\lambda^{+})
~.}}

\end{document}